\documentclass[graybox]{svmult}%
\usepackage{amsmath}
\usepackage{amsfonts}
\usepackage{amssymb}
\usepackage{graphicx}%
\providecommand{\U}[1]{\protect\rule{.1in}{.1in}}

\usepackage{mathptmx}       
\usepackage{helvet}         
\usepackage{courier}        
\usepackage{type1cm}        
\usepackage{makeidx}         
\usepackage{graphicx}        
\usepackage{multicol}        
\usepackage[bottom]{footmisc}

\makeindex             

\usepackage{epstopdf}
\usepackage{bm}
\usepackage{hyperref}
\DeclareGraphicsExtensions{.eps}
\allowdisplaybreaks
\hypersetup{
    bookmarks=true,         
    unicode=false,          
    pdftoolbar=true,        
    pdfmenubar=true,        
    pdffitwindow=false,     
    pdfstartview={FitH},    
    pdftitle={},    
    pdfauthor={},     
    pdfsubject={},   
    pdfcreator={Justin J. Yang},   
    pdfproducer={}, 
    pdfkeywords={}, 
    pdfnewwindow=true,      
    colorlinks=false,       
    linkcolor=red,          
    citecolor=green,        
    filecolor=magenta,      
    urlcolor=cyan           
} 
\begin{document}
\title*{Optimal shrinkage estimation in heteroscedastic hierarchical linear models}
\titlerunning{Optimal shrinkage estimation in heteroscedastic hierarchical linear models}
\author{S. C. Kou and Justin J. Yang}
\institute{S. C. Kou \at Department of Statistics, Harvard University\\
1 Oxford Street, 02138, Cambridge, MA, USA\\
\email{kou@stat.harvard.edu}
\and Justin J. Yang \at Department of Statistics, Harvard University\\
1 Oxford Street, 02138, Cambridge, MA, USA\\
\email{juchenjustinyang@fas.harvard.edu}}
%
%
\maketitle

\abstract*{Shrinkage estimators have profound impacts in statistics and in scientific
and engineering applications. In this article, we consider shrinkage estimation in
the presence of linear predictors. We formulate two heteroscedastic hierarchical
regression models and study optimal shrinkage estimators in each model. A
class of shrinkage estimators, both parametric and semiparametric, based on unbiased risk estimate (URE) is proposed
and is shown to be (asymptotically) optimal under mean squared error loss in each model.
Simulation study is conducted to compare the performance of the proposed methods
with existing shrinkage estimators. We also apply the method to real data and obtain
encouraging and interesting results.}

\abstract{Shrinkage estimators have profound impacts in statistics and in scientific
and engineering applications. In this article, we consider shrinkage estimation in
the presence of linear predictors. We formulate two heteroscedastic hierarchical
regression models and study optimal shrinkage estimators in each model. A
class of shrinkage estimators, both parametric and semiparametric, based on unbiased risk estimate (URE) is proposed
and is shown to be (asymptotically) optimal under mean squared error loss in each model.
Simulation study is conducted to compare the performance of the proposed methods
with existing shrinkage estimators. We also apply the method to real data and obtain
encouraging and interesting results.} 

\section{Introduction}

Shrinkage estimators, hierarchical models and empirical Bayes methods, dating
back to the groundbreaking works of \cite{Stein-1956} and \cite{Robbins-1956},
have profound impacts in statistics and in scientific and engineering
applications. They provide effective tools to pool information from
(scientifically) related populations for simultaneous inference---the data on
each population alone often do not lead to the most effective estimation, but
by pooling information from the related populations together (for example, by
shrinking toward their consensus \textquotedblleft center\textquotedblright),
one could often obtain more accurate estimate for each individual population.
Ever since the seminal works of \cite{Stein-1956} and \cite{James&Stein-1961},
an impressive list of articles has been devoted to the study of shrinkage
estimators in normal models, including
\cite{Stein-1962,Lindley-1962,Efron&Morris-1972,Efron&Morris-1973,Efron&Morris-1975,Berger&Strawderman-1996,Rubin-1980,Morris-1983,Green&Strawderman-1985,Jones-1991,Brown-2008}%
, among others.

In this article, we consider shrinkage estimation in the presence of linear
predictors. In particular, we study \emph{optimal} shrinkage estimators for
\emph{heteroscedastic} data under \emph{linear} models. Our study is motivated
by three main considerations. First, in many practical problems, one often
encounters heteroscedastic (unequal variance) data; for example, the sample
sizes for different groups are not all equal. Second, in many statistical
applications, in addition to the heteroscedastic response variable, one often
has predictors. For example, the predictors could represent longitudinal
patterns \cite{Fearn-1975,Hui&Berger-1983,Strenio&Weisberg&Bryk-1983}, exam
scores \cite{Rubin-1980}, characteristics of hospital patients
\cite{Normand&Glickman&Gatsonis-1997}, etc. Third, in applying shrinkage
estimators to real data, it is quite natural to ask for the \emph{optimal} way
of shrinkage.

The (risk) optimality is not addressed by the conventional estimators, such as
the empirical Bayes ones. One might wonder if such an optimal shrinkage
estimator exists in the first place. We shall see shortly that in fact
(asymptotically) optimal shrinkage estimators do exist and that the optimal
estimators are \emph{not} empirical Bayes ones but are characterized by an
unbiased risk estimate (URE).

The study of optimal shrinkage estimators under the heteroscedastic normal
model was first considered in \cite{Xie&Kou&Brown-2012}, where the
(asymptotic) optimal shrinkage estimator was identified for both the
parametric and semiparametric cases. \cite{Xie&Kou&Brown-2015} extends the
(asymptotic) optimal shrinkage estimators to exponential families and
heteroscedastic location-scale families. The current article can be viewed as
an extension of the idea of optimal shrinkage estimators to heteroscedastic
linear models.

We want to emphasize that this article works on a theoretical setting somewhat
different from \cite{Xie&Kou&Brown-2015} but can still cover its main results.
Our theoretical results show that the optimality of the proposed URE shrinkage
estimators does not rely on normality nor on the tail behavior of the sampling
distribution. What we require here are the symmetry and the existence of the
fourth moment for the standardized variable.

This article is organized as follows. We first formulate the heteroscedastic
linear models in Sec.
\ref{Sec.: Heteroscedastic Hierarchical Linear Models}. Interestingly, there
are two parallel ways to do so, and both are natural extensions of the
heteroscedastic normal model. After reviewing the conventional empirical Bayes
methods, we introduce the construction of our optimal shrinkage estimators for
heteroscedastic linear models in Sec. \ref{Sec.: URE Estimates}. The
optimal shrinkage estimators are based on an unbiased risk estimate (URE). We
show in Sec. \ref{Sec.: Theory of URE} that the URE shrinkage estimators
are asymptotically optimal in risk. In Sec. \ref{Sec.: Semiparametric URE}
we extend the shrinkage estimation to a semiparametric family. Simulation
studies are conducted in Sec. \ref{Sec.: Simulation}. We apply the URE
shrinkage estimators in Sec. \ref{Sec.: Baseball data analysis} to the
baseball data set of \cite{Brown-2008} and observe quite interesting and
encouraging results. We conclude in Sec. \ref{Sec.: Conclusion} with some
discussion and extension. The appendix details the proofs and derivations for
the theoretical results.

\section{Heteroscedastic Hierarchical Linear Models}

\label{Sec.: Heteroscedastic Hierarchical Linear Models}

Consider the heteroscedastic estimation problem%
\begin{equation}
Y_{i}|\boldsymbol{\theta}\overset{\text{indep.}}{\sim}\mathcal{N}\left(
\theta_{i},A_{i}\right)  ,\ \ \ \ i=1,...,p,
\label{Data Generating Distribution}%
\end{equation}
where $\boldsymbol{\theta}=\left(  \theta_{1},...,\theta_{p}\right)  ^{T}$ is
the unknown mean vector, which is to be estimated, and the variances $A_{i}>0$
are unequal, which are assumed to be known. In many statistical applications,
in addition to the heteroscedastic $\boldsymbol{Y}=\left(  Y_{1}%
,...,Y_{p}\right)  ^{T}$, one often has predictors $\boldsymbol{X}$. A natural
question is to consider a heteroscedastic linear model that incorporates these
covariates. Notation-wise, let $\{Y_{i},\boldsymbol{X}_{i}\}_{i=1}^{p}$ denote
the $p$ independent statistical units, where $Y_{i}$ is the response variable
of the $i$-th unit, and $\boldsymbol{X}_{i}=(X_{1i},\ldots,X_{ki})^{T}$ is a
$k$-dimensional column vector that corresponds to the $k$ covariates of the
$i$-th unit. The $k\times p$ matrix
\[
\boldsymbol{X}=\left[  \boldsymbol{X}_{1}|\cdots|\boldsymbol{X}_{p}\right]
,\ \ \ \ \boldsymbol{X}_{1},..,\boldsymbol{X}_{p}\in\mathbb{R}^{k},
\]
where $\boldsymbol{X}_{i}$ is the $i$-th column of $\boldsymbol{X}$, then
contains the covariates for all the units. Throughout this article we assume
that $\boldsymbol{X}$ has full rank, i.e., $\mathrm{rank}(\boldsymbol{X})=k$.

To include the predictors, we note that, interestingly, there are \emph{two}
different ways to build up a heteroscedastic hierarchical linear model, which
lead to different structure for shrinkage estimation.

\begin{figure}[t]
\centering\includegraphics[width=1\linewidth]{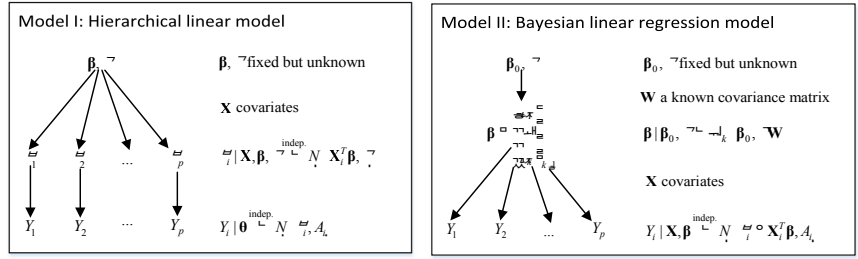}\caption{Graphical
illustration of the two heteroscedastic hierarchical linear models.}%
\label{fig.: Model illustration}%
\end{figure}

\begin{description}
\item[Model I: Hierarchical linear model.] On top of
(\ref{Data Generating Distribution}), the $\theta_{i}$'s are $\theta
_{i}\overset{\text{indep.}}{\sim}\mathcal{N}\left(  \boldsymbol{X}_{i}%
^{T}\boldsymbol{\beta},\lambda\right)  $, where $\boldsymbol{\beta}$ and
$\lambda$ are both \emph{unknown} hyper-parameters. Model I\ has been
suggested as early as \cite{Stein-1966}. See \cite{Morris-1983} and
\cite{Morris&Lysy-2012} for more discussions. The special case of no
covariates (i.e., $k=1$ and $\boldsymbol{X}=\left[  1|\cdots|1\right]  $) is
studied in depth in \cite{Xie&Kou&Brown-2012}.

\item[Model II: Bayesian linear regression model.] Together with
(\ref{Data Generating Distribution}), one assumes $\boldsymbol{\theta
}=\boldsymbol{X}^{T}\boldsymbol{\beta}$ with $\boldsymbol{\beta}$ following a
conjugate prior distribution $\boldsymbol{\beta}\sim\mathcal{N}_{k}\left(
\boldsymbol{\beta}_{0},\lambda\boldsymbol{W}\right)  $, where $\boldsymbol{W}$
is a \emph{known} $k\times k$ positive definite matrix and $\boldsymbol{\beta
}_{0}$ and $\lambda$ are \emph{unknown} hyper-parameters. Model II has been
considered in
\cite{Lindley&Smith-1972,Copas-1983,Raftery&Madigan&Hoeting-1997} among
others; it includes ridge regression as a special case when $\boldsymbol{\beta
}_{0}=\boldsymbol{0}_{k}$ and $\boldsymbol{W}=\boldsymbol{I}_{k}$.
\end{description}

Figure \ref{fig.: Model illustration} illustrates these two hierarchical
linear models. Under Model I, the posterior mean of $\boldsymbol{\theta}$ is
$\hat{\theta}_{i}^{\lambda,\boldsymbol{\beta}}=\lambda\left(  \lambda
+A_{i}\right)  ^{-1}Y_{i}+A_{i}\left(  \lambda+A_{i}\right)  ^{-1}%
\boldsymbol{X}_{i}^{T}\boldsymbol{\beta}$ for $i=1,...,p$, so the shrinkage
estimation is formed by directly shrinking the raw observation $Y_{i}$ toward
a linear combination of the $k$ covariates $\boldsymbol{X}_{i}$. If we denote
$\mu_{i}=\boldsymbol{X}_{i}^{T}\boldsymbol{\beta}$, and $\boldsymbol{\mu
}=\left(  \mu_{1},...,\mu_{p}\right)  ^{T}\in\mathcal{L}_{\mathrm{row}}\left(
\boldsymbol{X}\right)  $, the row space of $\boldsymbol{X}$, then we can
rewrite the posterior mean of $\boldsymbol{\theta}$ under Model I as%
\begin{equation}
\hat{\theta}^{\lambda,\boldsymbol{\mu}}=\dfrac{\lambda}{\lambda+A_{i}}%
Y_{i}+\dfrac{A_{i}}{\lambda+A_{i}}\mu_{i},\ \ \ \ \text{with }\boldsymbol{\mu
}\in\mathcal{L}_{\mathrm{row}}\left(  \boldsymbol{X}\right)  .
\label{Model I Shrinkage estimator}%
\end{equation}

Under Model II, the posterior mean of $\boldsymbol{\theta}$ is%
\begin{equation}
\boldsymbol{\hat{\theta}}^{\lambda,\boldsymbol{\beta}_{0}}=\boldsymbol{X}%
^{T}\boldsymbol{\hat{\beta}}^{\lambda,\boldsymbol{\beta}_{0}},\ \ \ \text{with
}\boldsymbol{\hat{\beta}}^{\lambda,\boldsymbol{\beta}_{0}}=\lambda
\boldsymbol{W}(\lambda\boldsymbol{W}+\boldsymbol{V})^{-1}\boldsymbol{\hat
{\beta}}^{\mathrm{WLS}}+\boldsymbol{V}\left(  \lambda\boldsymbol{W}%
+\boldsymbol{V}\right)  ^{-1}\boldsymbol{\beta}_{0},
\label{Model II Shrinkage estimator}%
\end{equation}
where $\boldsymbol{\hat{\beta}}^{\mathrm{WLS}}=\left(  \boldsymbol{XA}%
^{-1}\boldsymbol{X}^{T}\right)  ^{-1}\boldsymbol{XA}^{-1}\boldsymbol{Y}$ is
the weighted least squares estimate of the regression coefficient,
$\boldsymbol{A}$ is the diagonal matrix $\boldsymbol{A}=\mathrm{diag}\left(
A_{1},...,A_{p}\right)  $, and $\boldsymbol{V}=(\boldsymbol{XA}^{-1}%
\boldsymbol{X}^{T})^{-1}$. Thus, the estimate for $\theta_{i}$ is linear in
$\boldsymbol{X}_{i}$, and the \textquotedblleft shrinkage\textquotedblright%
\ is achieved by shrinking the regression coefficient from the weighted least
squares estimate $\boldsymbol{\hat{\beta}}^{\mathrm{WLS}}$ toward the prior
coefficient $\boldsymbol{\beta}_{0}$.

As both Models I and II are natural generalizations of the heteroscedastic
normal model (\ref{Data Generating Distribution}), we want\ to investigate if
there is an optimal choice of the hyper-parameters in each case. Specifically,
we want to investigate the best empirical choice of the hyper-parameters in
each case under the mean squared error loss%
\begin{equation}
l_{p}(\boldsymbol{\theta},\boldsymbol{\hat{\theta}})=\dfrac{1}{p}\left\Vert
\boldsymbol{\theta}-\boldsymbol{\hat{\theta}}\right\Vert ^{2}=\dfrac{1}{p}%
\sum_{i=1}^{p}\left(  \theta_{i}-\hat{\theta}_{i}\right)  ^{2}
\label{Loss Function}%
\end{equation}
with the associated risk of $\boldsymbol{\hat{\theta}}$ defined by%
\[
R_{p}(\boldsymbol{\theta},\boldsymbol{\hat{\theta}})=\mathbb{E}%
_{\boldsymbol{Y}|\boldsymbol{\theta}}\left(  l_{p}(\boldsymbol{\theta
},\boldsymbol{\hat{\theta}})\right)  ,
\]
where the expectation is taken with respect to $\boldsymbol{Y}$ given
$\boldsymbol{\theta}$.

\begin{remark}
Even though we start from the Bayesian setting to motivate the form of
shrinkage estimators, our discussion will be all based on the frequentist
setting. Hence all probabilities and expectations throughout this article are
fixed at the unknown true $\boldsymbol{\theta}$.
\end{remark}

\begin{remark}
The diagonal assumption of $\boldsymbol{A}$ is quite important for Model I but
not so for Model II, as in Model II we can always apply some linear
transformations to obtain a diagonal covariance matrix. Without loss of
generality, we will keep the diagonal assumption for $\boldsymbol{A}$ in Model II.
\end{remark}

For the ease of exposition, we will next overview the conventional empirical
Bayes estimates in a general two-level hierarchical model, which includes both
Models I and II:%
\begin{equation}
\boldsymbol{Y}|\boldsymbol{\theta}\sim\mathcal{N}_{p}(\boldsymbol{\theta
},\boldsymbol{A})\text{ and }\boldsymbol{\theta}\sim\mathcal{N}_{p}%
(\boldsymbol{\mu},\boldsymbol{B}),
\label{Generic two level hierarchical regression model}%
\end{equation}
where $\boldsymbol{B}$ is a non-negative definite symmetric matrix that is
restricted in an allowable set $\mathcal{B}$, and $\boldsymbol{\mu}$ is in the
row space $\mathcal{L}_{\mathrm{row}}(\boldsymbol{X})$ of $\boldsymbol{X}$.

\begin{remark}
Under Model I, $\boldsymbol{\mu}$ and $\boldsymbol{B}$ take the form of
$\boldsymbol{\mu}=\boldsymbol{X}^{T}\boldsymbol{\beta}$ and $\boldsymbol{B}%
\in\mathcal{B}=\left\{  \lambda\boldsymbol{I}_{p}:\lambda>0\right\}  $,
whereas under Model II, $\boldsymbol{\mu}$ and $\boldsymbol{B}$ take the form
of $\boldsymbol{\mu}=\boldsymbol{X}^{T}\boldsymbol{\beta}_{0}$ and
$\boldsymbol{B}\in\mathcal{B}=\left\{  \lambda\boldsymbol{X}^{T}%
\boldsymbol{WX}:\lambda>0\right\}  $. It is interesting to observe that in
Model I, $\boldsymbol{B}$ is of full rank, while in Model II, $\boldsymbol{B}$
is of rank $k$. As we shall see, this distinction will have interesting
theoretical implications for the optimal shrinkage estimators.
\end{remark}

\begin{lemma}
\label{Lma.: Generic posterior dist. and marginal}Under the two-level
hierarchical model (\ref{Generic two level hierarchical regression model}),
the posterior distribution is%
\[
\boldsymbol{\theta}|\boldsymbol{Y}\sim\mathcal{N}_{p}\left(  \boldsymbol{B}%
(\boldsymbol{A}+\boldsymbol{B})^{-1}\boldsymbol{Y}+\boldsymbol{A}%
(\boldsymbol{A}+\boldsymbol{B})^{-1}\boldsymbol{\mu},\boldsymbol{A}%
(\boldsymbol{A}+\boldsymbol{B})^{-1}\boldsymbol{B}\right)  ,
\]
and the marginal distribution of $\boldsymbol{Y}$ is $\boldsymbol{Y}%
\sim\mathcal{N}_{p}\left(  \boldsymbol{\mu}, \boldsymbol{A}+\boldsymbol{B}%
\right) $.
\end{lemma}

For given values of $\boldsymbol{B}$ and $\boldsymbol{\mu}$, the posterior
mean of the parameter $\boldsymbol{\theta}$ leads to the Bayes estimate%
\begin{equation}
\boldsymbol{\hat{\theta}}^{\boldsymbol{B},\boldsymbol{\mu}}=\boldsymbol{B}%
(\boldsymbol{A}+\boldsymbol{B})^{-1}\boldsymbol{Y}+\boldsymbol{A}%
(\boldsymbol{A}+\boldsymbol{B})^{-1}\boldsymbol{\mu}.
\label{Generic Shrinkage Estimator}%
\end{equation}
To use the Bayes estimate in practice, one has to specify the hyper-parameters
in $\boldsymbol{B}$ and $\boldsymbol{\mu}$. The conventional empirical Bayes
method uses the marginal distribution of $\boldsymbol{Y}$ to estimate the
hyper-parameters. For instance, the empirical Bayes maximum likelihood
estimates (EBMLE) $\boldsymbol{\hat{B}}^{\mathrm{EBMLE}}$ and
$\boldsymbol{\hat{\mu}}^{\mathrm{EBMLE}}$ are obtained by maximizing the
marginal likelihood of $\boldsymbol{Y}$:%
\[
\left(  \boldsymbol{\hat{B}}^{\mathrm{EBMLE}},\boldsymbol{\hat{\mu}%
}^{\mathrm{EBMLE}}\right)  =\operatorname*{argmax}%
\limits_{\substack{\boldsymbol{B}\in\mathcal{B}\\\boldsymbol{\mu}%
\in\mathcal{L}_{\mathrm{row}}\left(  \boldsymbol{X}\right)  }}-\left(
\boldsymbol{Y}-\boldsymbol{\mu}\right)  ^{T}\left(  \boldsymbol{A}%
+\boldsymbol{B}\right)  ^{-1}\left(  \boldsymbol{Y}-\boldsymbol{\mu}\right)
-\log\left(  \det\left(  \boldsymbol{A}+\boldsymbol{B}\right)  \right)  .
\]
Alternatively, the empirical Bayes method-of-moment estimates (EBMOM)
$\boldsymbol{\hat{B}}^{\mathrm{EBMOM}}$ and $\boldsymbol{\hat{\mu}%
}^{\mathrm{EBMOM}}$ are obtained by solving the following moment equations for
$\boldsymbol{B}\in\mathcal{B}$ and $\boldsymbol{\mu}\in\mathcal{L}%
_{\mathrm{row}}\left(  \boldsymbol{X}\right)  $:%
\begin{align*}
\boldsymbol{\mu}  &  =\boldsymbol{X}^{T}\left(  \boldsymbol{X}\left(
\boldsymbol{A}+\boldsymbol{B}\right)  ^{-1}\boldsymbol{X}^{T}\right)
^{-1}\boldsymbol{X}\left(  \boldsymbol{A}+\boldsymbol{B}\right)
^{-1}\boldsymbol{Y},\\
\boldsymbol{B}  &  =\left(  \boldsymbol{Y}-\boldsymbol{\mu}\right)  \left(
\boldsymbol{Y}-\boldsymbol{\mu}\right)  ^{T}-\boldsymbol{A}.
\end{align*}
If no solutions of $\boldsymbol{B}$ can be found in $\mathcal{B}$, we then set
$\boldsymbol{\hat{B}}^{\mathrm{EBMOM}}=\boldsymbol{0}_{p\times p}$. Adjustment
for the loss of $k$ degrees of freedom from the estimation of $\boldsymbol{\mu
}$ might be applicable for $\boldsymbol{B}=\lambda\boldsymbol{C}$
($\boldsymbol{C}=\boldsymbol{I}_{p}$ for Model I and $\boldsymbol{X}%
^{T}\boldsymbol{WX}$ for Model II): we can replace the second moment equation
by%
\[
\lambda=\left(  \dfrac{p}{p-k}\dfrac{\left\Vert \boldsymbol{Y}-\boldsymbol{\mu
}\right\Vert ^{2}}{\mathrm{tr}\left(  \boldsymbol{C}\right)  }-\dfrac
{\mathrm{tr}\left(  \boldsymbol{A}\right)  }{\mathrm{tr}\left(  \boldsymbol{C}%
\right)  }\right)  ^{+}.
\]
The corresponding empirical Bayes shrinkage estimator $\boldsymbol{\hat
{\theta}}^{\mathrm{EBMLE}}$ or $\boldsymbol{\hat{\theta}}^{\mathrm{EBMOM}}$ is
then formed by plugging $(\boldsymbol{\hat{B}}^{\mathrm{EBMLE}}%
,\boldsymbol{\hat{\mu}}^{\mathrm{EBMLE}})$ or $(\boldsymbol{\hat{B}%
}^{\mathrm{EBMOM}},\boldsymbol{\hat{\mu}}^{\mathrm{EBMOM}})$ into equation
(\ref{Generic Shrinkage Estimator}).

\section{URE Estimates}

\label{Sec.: URE Estimates}

The formulation of the empirical Bayes estimates raises a natural question:
which one is preferred $\boldsymbol{\hat{\theta}}^{\mathrm{EBMLE}}$ or
$\boldsymbol{\hat{\theta}}^{\mathrm{EBMOM}}$? More generally, is there an
optimal way to choose the hyper-parameters? It turns out that neither
$\boldsymbol{\hat{\theta}}^{\mathrm{EBMLE}}$ nor $\boldsymbol{\hat{\theta}%
}^{\mathrm{EBMOM}}$ is optimal. The (asymptotically) optimal estimate, instead
of relying on the marginal distribution of $\boldsymbol{Y}$, is characterized
by an unbiased risk estimate (URE). The idea of forming a shrinkage estimate
through URE for heteroscedastic models is first suggested in
\cite{Xie&Kou&Brown-2012}. We shall see that in our context of hierarchical
linear models (both Models I\ and II) the URE estimators that we are about to
introduce have (asymptotically) optimal risk properties.

The basic idea behind URE estimators is the following. Ideally we want to find
the hyper-parameters that give the smallest risk. However, since the risk
function depends on the unknown $\boldsymbol{\theta}$, we cannot directly
minimize the risk function in practice. If we can find a good estimate of the
risk function instead, then minimizing this proxy of the risk will lead to a
competitive estimator.

To formally introduce the URE estimators, we start from the observation that,
under the mean squared error loss (\ref{Loss Function}), the risk of the Bayes
estimator $\boldsymbol{\hat{\theta}}^{\boldsymbol{B},\boldsymbol{\mu}}$ for
fixed $\boldsymbol{B}$ and $\boldsymbol{\mu}$ is%
\begin{equation}
R_{p}(\boldsymbol{\theta},\boldsymbol{\hat{\theta}}^{\boldsymbol{B}%
,\boldsymbol{\mu}})=\dfrac{1}{p}\left\Vert \boldsymbol{A}\left(
\boldsymbol{A}+\boldsymbol{B}\right)  ^{-1}\left(  \boldsymbol{\mu
}-\boldsymbol{\theta}\right)  \right\Vert ^{2}+\dfrac{1}{p}\mathrm{tr}\left(
\boldsymbol{B}\left(  \boldsymbol{A}+\boldsymbol{B}\right)  ^{-1}%
\boldsymbol{A}\left(  \boldsymbol{A}+\boldsymbol{B}\right)  ^{-1}%
\boldsymbol{B}\right)  , \label{Generic Risk}%
\end{equation}
which can be easily shown using the bias-variance decomposition of the mean
squared error. As the risk function involves the unknown $\boldsymbol{\theta}%
$, we cannot directly minimize it. However, an unbiased estimate of the risk
is available:%
\begin{equation}
\mathrm{URE}\left(  \boldsymbol{B},\boldsymbol{\mu}\right)  =\dfrac{1}%
{p}\left\Vert \boldsymbol{A}\left(  \boldsymbol{A}+\boldsymbol{B}\right)
^{-1}\left(  \boldsymbol{Y}-\boldsymbol{\mu}\right)  \right\Vert ^{2}%
+\dfrac{1}{p}\mathrm{tr}\left(  \boldsymbol{A}-2\boldsymbol{A}\left(
\boldsymbol{A}+\boldsymbol{B}\right)  ^{-1}\boldsymbol{A}\right)  ,
\label{Generic URE}%
\end{equation}
which again can be easily shown using the bias-variance decomposition of the
mean squared error. Intuitively, if $\mathrm{URE}\left(  \boldsymbol{B}%
,\boldsymbol{\mu}\right)  $ is a good approximation of the actual risk, then
we would expect the estimator obtained by minimizing the URE to have good
properties. This leads to the URE estimator $\boldsymbol{\hat{\theta}%
}^{\mathrm{URE}}$, defined by%
\begin{equation}
\boldsymbol{\hat{\theta}}^{\mathrm{URE}}=\boldsymbol{\hat{B}}^{\mathrm{URE}%
}(\boldsymbol{A}+\boldsymbol{\hat{B}}^{\mathrm{URE}})^{-1}\boldsymbol{Y}%
+\boldsymbol{A}(\boldsymbol{A}+\boldsymbol{\hat{B}}^{\mathrm{URE}}%
)^{-1}\boldsymbol{\hat{\mu}}^{\mathrm{URE}}\mathbf{,} \label{genericUREest}%
\end{equation}
where%
\[
\left(  \boldsymbol{\hat{B}}^{\mathrm{URE}},\boldsymbol{\hat{\mu}%
}^{\mathrm{URE}}\right)  =\operatorname*{argmin}\limits_{B\in\mathcal{B}%
,\ \boldsymbol{\mu}\in\mathcal{L}_{\mathrm{row}}\left(  \boldsymbol{X}\right)
}\mathrm{URE}\left(  \boldsymbol{B},\boldsymbol{\mu}\right)  .
\]

In the URE estimator (\ref{genericUREest}), $\boldsymbol{\hat{B}%
}^{\mathrm{URE}}$ and $\boldsymbol{\hat{\mu}}^{\mathrm{URE}}$ are jointly
determined by minimizing the URE. When the number of independent statistical
units $p$ is small or moderate, joint minimization of $\boldsymbol{B}$ and the
vector $\boldsymbol{\mu}$, however, may be too ambitious. In this setting, it
might be beneficial to set $\boldsymbol{\mu}$ by a predetermined rule and only
optimize $\boldsymbol{B}$, as it might reduce the variability of the resulting
estimate. In particular, we can consider shrinking toward a generalized least
squares (GLS) regression estimate
\[
\boldsymbol{\hat{\mu}}^{\boldsymbol{M}}=\boldsymbol{X}^{T}\left(
\boldsymbol{XMX}^{T}\right)  ^{-1}\boldsymbol{XMY}=\boldsymbol{P}%
_{\boldsymbol{M},\boldsymbol{X}}\boldsymbol{Y},
\]
where $\boldsymbol{M}$ is a \emph{prespecified} symmetric positive definite
matrix. This use of $\boldsymbol{\hat{\mu}}^{\boldsymbol{M}}$ gives the
shrinkage estimate $\boldsymbol{\hat{\theta}}^{\boldsymbol{B},\boldsymbol{\hat
{\mu}}^{\boldsymbol{M}}}=\boldsymbol{B}(\boldsymbol{A}+\boldsymbol{B}%
)^{-1}\boldsymbol{Y}+\boldsymbol{A}(\boldsymbol{A}+\boldsymbol{B}%
)^{-1}\boldsymbol{\hat{\mu}}^{\boldsymbol{M}}$, where one only needs to
determine $\boldsymbol{B}$. We can construct another URE estimate for this
purpose. Similar to the previous construction, we note that $\boldsymbol{\hat
{\theta}}^{\boldsymbol{B},\boldsymbol{\hat{\mu}}^{\boldsymbol{M}}}$ has risk%
\begin{align}
&  R_{p}(\boldsymbol{\theta},\boldsymbol{\hat{\theta}}^{\boldsymbol{B}%
,\boldsymbol{\hat{\mu}}^{\boldsymbol{M}}})=\dfrac{1}{p}\left\Vert
\boldsymbol{A}\left(  \boldsymbol{A}+\boldsymbol{B}\right)  ^{-1}\left(
\boldsymbol{I}_{p}-\boldsymbol{P}_{\boldsymbol{M},\boldsymbol{X}}\right)
\boldsymbol{\theta}\right\Vert ^{2}\nonumber\\
&  +\dfrac{1}{p}\mathrm{tr}\left(  \left(  \boldsymbol{I}_{p}-\boldsymbol{A}%
\left(  \boldsymbol{A}+\boldsymbol{B}\right)  ^{-1}\left(  \boldsymbol{I}%
_{p}-\boldsymbol{P}_{\boldsymbol{M},\boldsymbol{X}}\right)  \right)
\boldsymbol{A}\left(  \boldsymbol{I}_{p}-\boldsymbol{A}\left(  \boldsymbol{A}%
+\boldsymbol{B}\right)  ^{-1}\left(  \boldsymbol{I}_{p}-\boldsymbol{P}%
_{\boldsymbol{M},\boldsymbol{X}}\right)  \right)  ^{T}\right)  . \label{risk2}%
\end{align}
An unbiased risk estimate of it is
\begin{equation}
\mathrm{URE}_{\boldsymbol{M}}\left(  \boldsymbol{B}\right)  =\dfrac{1}%
{p}\left\Vert \boldsymbol{A}\left(  \boldsymbol{A}+\boldsymbol{B}\right)
^{-1}\left(  \boldsymbol{Y}-\boldsymbol{\hat{\mu}}^{\boldsymbol{M}}\right)
\right\Vert ^{2}+\dfrac{1}{p}\mathrm{tr}\left(  \boldsymbol{A}-2\boldsymbol{A}%
\left(  \boldsymbol{A}+\boldsymbol{B}\right)  ^{-1}\left(  \boldsymbol{I}%
_{p}-\boldsymbol{P}_{\boldsymbol{M},\boldsymbol{X}}\right)  \boldsymbol{A}%
\right)  . \label{URE2}%
\end{equation}
Both (\ref{risk2}) and (\ref{URE2}) can be easily proved by the bias-variance
decomposition of mean squared error. Minimizing $\mathrm{URE}_{\boldsymbol{M}%
}\left(  \boldsymbol{B}\right)  $ over $\boldsymbol{B}$ gives the URE GLS
shrinkage estimator (which shrinks toward $\boldsymbol{\hat{\mu}%
}^{\boldsymbol{M}}$):%
\begin{equation}
\boldsymbol{\hat{\theta}}_{\boldsymbol{M}}^{\mathrm{URE}}=\boldsymbol{\hat{B}%
}_{\boldsymbol{M}}^{\mathrm{URE}}\left(  \boldsymbol{A}+\boldsymbol{\hat{B}%
}_{\boldsymbol{M}}^{\mathrm{URE}}\right)  ^{-1}\boldsymbol{Y}+\boldsymbol{A}%
\left(  \boldsymbol{A}+\boldsymbol{\hat{B}}_{\boldsymbol{M}}^{\mathrm{URE}%
}\right)  ^{-1}\boldsymbol{\hat{\mu}}^{\boldsymbol{M}}\mathbf{,}
\label{UREMEst}%
\end{equation}
where%
\[
\boldsymbol{\hat{B}}_{\boldsymbol{M}}^{\mathrm{URE}}\boldsymbol{=}%
\operatorname*{argmin}\limits_{\boldsymbol{B}\in\mathcal{B}}\mathrm{URE}%
_{\boldsymbol{M}}\left(  \boldsymbol{B}\right)  .
\]

\begin{remark}
When $\boldsymbol{M}=\boldsymbol{I}_{p}$, clearly $\boldsymbol{\hat{\mu}%
}^{\boldsymbol{M}}=\boldsymbol{\hat{\mu}}^{\mathrm{OLS}}$, the ordinary least squares regression estimate. When $\boldsymbol{M}=\boldsymbol{A}^{-1}$, then
$\boldsymbol{\hat{\mu}}^{\boldsymbol{M}}=\boldsymbol{\hat{\mu}}^{\mathrm{WLS}%
}$, the weighted least squares regression estimate.
\end{remark}

\section{Theoretical Properties of URE Estimates}

\label{Sec.: Theory of URE}

This section is devoted to the risk properties of the URE estimators. Our core
theoretical result is to show that the risk estimate URE is not only unbiased
for the risk but, more importantly, uniformly close to the actual loss. We
therefore expect that minimizing URE would lead to an estimate with
competitive risk properties.

\subsection{Uniform Convergence of URE}

To present our theoretical result, we first define $\mathcal{L}$ to be a
subset of $\mathcal{L}_{\mathrm{row}}\left(  \boldsymbol{X}\right)  $:%
\[
\mathcal{L}=\{\boldsymbol{\mu}\in\mathcal{L}_{\mathrm{row}}\left(
\boldsymbol{X}\right)  :\left\Vert \boldsymbol{\mu}\right\Vert \leq
Mp^{\kappa}\left\Vert \boldsymbol{Y}\right\Vert \},
\]
where $M$ is a large and fixed constant and $\kappa\in\lbrack0,1/2)$ is a
constant. Next, we introduce the following regularity conditions:

(A) $\sum_{i=1}^{p}A_{i}^{2}=O\left(  p\right)  $; (B) $\sum_{i=1}^{p}%
A_{i}\theta_{i}^{2}=O\left(  p\right)  $; (C) $\sum_{i=1}^{p}\theta_{i}%
^{2}=O\left(  p\right)  $;

(D) $p^{-1}\boldsymbol{XAX}^{T}\rightarrow\boldsymbol{\Omega}_{D}$; (E)
$p^{-1}\boldsymbol{XX}^{T}\rightarrow\boldsymbol{\Omega}_{\boldsymbol{E}}>0$;

(F) $p^{-1}\boldsymbol{XA}^{-1}\boldsymbol{X}^{T}\rightarrow\boldsymbol{\Omega
}_{F}>0$; (G) $p^{-1}\boldsymbol{XA}^{-2}\boldsymbol{X}^{T}\rightarrow
\boldsymbol{\Omega}_{G}$.

The theorem below shows that $\mathrm{URE}\left(  \boldsymbol{B}%
,\boldsymbol{\mu}\right)  $ not only unbiasedly estimates the risk but also is
(asymptotically) uniformly close to the actual loss.

\begin{theorem}
\label{Thm.: Main Theorem} Assume conditions (A)-(E) for Model I or assume
conditions (A) and (D)-(G) for Model II. In either case, we have%
\[
\sup\limits_{\boldsymbol{B}\in\mathcal{B},\ \boldsymbol{\mu}\in\mathcal{L}%
}\left\vert \mathrm{URE}\left(  \boldsymbol{B},\boldsymbol{\mu}\right)
-l_{p}\left(  \boldsymbol{\theta},\boldsymbol{\hat{\theta}}^{\boldsymbol{B}%
,\boldsymbol{\mu}}\right)  \right\vert \rightarrow0\text{ in }L^{1},\text{ as
}p\rightarrow\infty.
\]
\end{theorem}

We want to remark here that the set $\mathcal{L}$ gives the allowable range of
$\boldsymbol{\mu}$: the norm of $\boldsymbol{\mu}$ is up to an $o\left(
p^{1/2}\right)  $ multiple of the norm of $\boldsymbol{Y}$. This choice of
$\mathcal{L}$ does not lead to any difficulty in practice because, given a
large enough constant $M$, it will cover the shrinkage locations of any
sensible shrinkage estimator. We note that it is possible to define the range
of sensible shrinkage locations in other ways (e.g., one might want to define
it by $\infty$-norm in $\mathbb{R}^{p}$), but we find our setting more
theoretically appealing and easy to work with. In particular, our assumption
of the exponent $\kappa<1/2$ is flexible enough to cover most interesting
cases, including $\boldsymbol{\hat{\mu}}^{\mathrm{OLS}}$, the ordinary least
squares regression estimate, and $\boldsymbol{\hat{\mu}}^{\mathrm{WLS}}$, the
weighted least squares regression estimate (as in Remark 4) as shown in the
following lemma.

\begin{lemma}
\label{Lma.: WLS is in the restriction class} (i) $\boldsymbol{\hat{\mu}%
}^{\mathrm{OLS}}\in\mathcal{L}$. (ii) Assume $\left(  \mathrm{A}\right)  $ and
$\left(  \mathrm{A}^{\prime}\right)  \ \sum_{i=1}^{p}A_{i}^{-2-\delta
}=O\left(  p\right)  $ for some $\delta>0$; then $\boldsymbol{\hat{\mu}%
}^{\mathrm{WLS}}\in\mathcal{L}$ for $\kappa=4^{-1}+\left( 4+2 \delta \right) ^{-1}$
and a large enough $M$.
\end{lemma}

\begin{remark}
We want to mention here that Theorem \ref{Thm.: Main Theorem} in the case of
Model I covers Theorem 5.1 of \cite{Xie&Kou&Brown-2012} (which is the special
case of $k=1$ and $\boldsymbol{X}=\left[  1|1|...|1\right]  $) because the
restriction of $\left\vert \mu\right\vert \leq\max\limits_{1\leq i\leq
p}\left\vert Y_{i}\right\vert $ in \cite{Xie&Kou&Brown-2012} is contained in
$\mathcal{L}$ as%
\[
\max\limits_{1\leq i\leq p}\left\vert Y_{i}\right\vert = ( \max\limits_{1\leq
i\leq p}Y_{i}^{2} ) ^{1/2}\leq ( \sum_{i=1}^{p}Y_{i}^{2} ) ^{1/2}=\left\Vert
\boldsymbol{Y}\right\Vert .
\]
Furthermore, we do not require the stronger assumption of $\sum_{i=1}%
^{p}\left\vert \theta_{i}\right\vert ^{2+\delta}=O\left(  p\right)  $ for some
$\delta>0$ made in \cite{Xie&Kou&Brown-2012}. Note that in this case ($k=1$
and $\boldsymbol{X}=\left[  1|1|...|1\right]  $) we do not even require
conditions $\left(  \mathrm{D}\right)  $ and $\left(  \mathrm{E}\right)  $,
as condition $\left(  \mathrm{A}\right)  $ directly implies $\mathrm{tr} ( \left( \boldsymbol{X}\boldsymbol{X}^{T}\right) ^{-1} \boldsymbol{X}\boldsymbol{A}\boldsymbol{X}^{T} ) = O\left( 1\right)$,
the result we need in the proof of Theorem \ref{Thm.: Main Theorem} for Model I.
\end{remark}

\begin{remark}
In the proof of Theorem \ref{Thm.: Main Theorem}, the sampling distribution of
$\boldsymbol{Y}$ is involved only through the moment calculations, such as
$\mathbb{E} ( \mathrm{tr}( \boldsymbol{YY}^{T}-\boldsymbol{A}%
-\boldsymbol{\theta\theta}^{T} ) ^{2} ) $ and $\mathbb{E} ( \left\Vert
\boldsymbol{Y}\right\Vert ^{2} ) $. It is therefore straightforward to
generalize Theorem \ref{Thm.: Main Theorem} to the case of
\[
Y_{i}=\theta_{i}+\sqrt{A_{i}}Z_{i},
\]
where $Z_{i}$ follows \emph{any} distribution with mean $0$, variance $1$,
$\mathbb{E}\left(  Z_{i}^{3}\right)  =0$, and $\mathbb{E}\left(  Z_{i}%
^{4}\right)  <\infty$. This is noteworthy as our result also covers that of
\cite{Xie&Kou&Brown-2015} but the methodology we employ here does not require
to control the tail behavior of $Z_{i}$ as in
\cite{Xie&Kou&Brown-2012,Xie&Kou&Brown-2015}.
\end{remark}

\subsection{Risk Optimality}

In this section, we consider the risk properties of the URE estimators. We
will show that, under the hierarchical linear models, the URE estimators have
(asymptotically) optimal risk, whereas it is not necessarily so for other
shrinkage estimators such as the empirical Bayes ones.

A direct consequence of the uniform convergence of URE is that the URE
estimator has a loss/risk that is asymptotically no larger than that of any
other shrinkage estimators. Furthermore, the URE estimator is asymptotically
as good as the oracle loss estimator. To be precise, let $\boldsymbol{\tilde
{\theta}}^{\mathrm{OL}}$ be the oracle loss (OL) estimator defined by plugging%
\begin{align*}
\left(  \boldsymbol{\tilde{B}}^{\mathrm{OL}},\boldsymbol{\tilde{\mu}%
}^{\mathrm{OL}}\right)   &  =\operatorname*{argmin}\limits_{B\in
\mathcal{B},\ \boldsymbol{\mu}\in\mathcal{L}}l_{p}\left(  \boldsymbol{\theta
},\boldsymbol{\hat{\theta}}^{\boldsymbol{B},\boldsymbol{\mu}}\right) \\
&  =\operatorname*{argmin}\limits_{B\in\mathcal{B},\ \boldsymbol{\mu}%
\in\mathcal{L}}\left\Vert \boldsymbol{B}(\boldsymbol{A}+\boldsymbol{B}%
)^{-1}\boldsymbol{Y}+\boldsymbol{A}(\boldsymbol{A}+\boldsymbol{B}%
)^{-1}\boldsymbol{\mu}-\boldsymbol{\theta}\right\Vert ^{2}%
\end{align*}
into (\ref{Generic Shrinkage Estimator}). Of course, $\boldsymbol{\tilde
{\theta}}^{\mathrm{OL}}$ is not really an estimator, since it depends on the
unknown $\boldsymbol{\theta}$ (hence we use the notation $\boldsymbol{\tilde
{\theta}}^{\mathrm{OL}}$ rather than $\boldsymbol{\hat{\theta}}^{\mathrm{OL}}%
$). Although not obtainable in practice, $\boldsymbol{\tilde{\theta}%
}^{\mathrm{OL}}$ lays down the theoretical limit that one can ever hope to
reach. The next theorem shows that the URE estimator $\boldsymbol{\hat{\theta
}}^{\mathrm{URE}}$ is asymptotically as good as the oracle loss estimator,
and, consequently, it is asymptotically at least as good as any other
shrinkage estimator.

\begin{theorem}
\label{Thm.: OL Risk Optimality}Assume the conditions of Theorem
\ref{Thm.: Main Theorem} and that $\boldsymbol{\hat{\mu}}^{\mathrm{URE}}%
\in\mathcal{L}$. Then%
\begin{align*}
\lim\limits_{p\rightarrow\infty}\mathbb{P}\left(  l_{p}\left(
\boldsymbol{\theta},\boldsymbol{\hat{\theta}}^{\mathrm{URE}}\right)  \geq
l_{p}\left(  \boldsymbol{\theta},\boldsymbol{\tilde{\theta}}^{\mathrm{OL}%
}\right)  +\epsilon\right)   &  =0\ \ \ \ \forall\epsilon>0,\\
\limsup\limits_{p\rightarrow\infty}\left(  R_{p}\left(  \boldsymbol{\theta
},\boldsymbol{\hat{\theta}}^{\mathrm{URE}}\right)  -R_{p}\left(
\boldsymbol{\theta},\boldsymbol{\tilde{\theta}}^{\mathrm{OL}}\right)  \right)
&  =0.
\end{align*}
\end{theorem}

\begin{corollary}
\label{Cor.: Risk properties} Assume the conditions of Theorem
\ref{Thm.: Main Theorem} and that $\boldsymbol{\hat{\mu}}^{\mathrm{URE}}%
\in\mathcal{L}$. Then for any estimator $\boldsymbol{\hat{\theta}%
}^{\boldsymbol{\hat{B}}_{p},\boldsymbol{\hat{\mu}}_{p}}=\boldsymbol{\hat{B}%
}_{p}\left( \boldsymbol{A}+\boldsymbol{\hat{B}}_{p} \right) ^{-1}\boldsymbol{Y}%
+\boldsymbol{A}\left( \boldsymbol{A}+\boldsymbol{\hat{B}}_{p} \right) ^{-1}\boldsymbol{\hat
{\mu}}_{p}$ with $\boldsymbol{\hat{B}}_{p}\in\mathcal{B}$ and
$\boldsymbol{\hat{\mu}}_{p}\in\mathcal{L}$, we always have%
\begin{align*}
\lim\limits_{p\rightarrow\infty}\mathbb{P}\left(  l_{p}\left(
\boldsymbol{\theta},\boldsymbol{\hat{\theta}}^{\mathrm{URE}}\right)  \geq
l_{p}\left(  \boldsymbol{\theta},\boldsymbol{\hat{\theta}}^{\boldsymbol{\hat
{B}}_{p},\boldsymbol{\hat{\mu}}_{p}}\right)  +\epsilon\right)   &
=0\ \ \ \ \forall\epsilon>0,\\
\limsup\limits_{p\rightarrow\infty}\left(  R_{p}\left(  \boldsymbol{\theta
},\boldsymbol{\hat{\theta}}^{\mathrm{URE}}\right)  -R_{p}\left(
\boldsymbol{\theta},\boldsymbol{\hat{\theta}}^{\boldsymbol{\hat{B}}%
_{p},\boldsymbol{\hat{\mu}}_{p}}\right)  \right)   &  \leq0.
\end{align*}
\end{corollary}

Corollary \ref{Cor.: Risk properties} tells us that the URE estimator in
either Model I or II is asymptotically optimal: it has (asymptotically) the
smallest loss and risk among all shrinkage estimators of the form
(\ref{Generic Shrinkage Estimator}).

\subsection{Shrinkage toward the Generalized Least Squares Estimate}

The risk optimality also holds when we consider the URE estimator
$\boldsymbol{\hat{\theta}}_{\boldsymbol{M}}^{\mathrm{URE}}$ that shrinks
toward the GLS regression estimate $\boldsymbol{\hat{\mu}}^{\boldsymbol{M}%
}=\boldsymbol{P}_{\boldsymbol{M},\boldsymbol{X}}\boldsymbol{Y}$ as introduced
in Sec.~\ref{Sec.: URE Estimates}.

\begin{theorem}
\label{Thm.: Main Theorem for shrinking toward GLS} Assume the conditions of
Theorem \ref{Thm.: Main Theorem}, $\boldsymbol{\hat{\mu}}^{\boldsymbol{M}}%
\in\mathcal{L}$, and
\begin{equation}
p^{-1}\boldsymbol{XMX}^{T}\rightarrow\boldsymbol{\Omega}_{1}>0,\ \ \ \ %
p^{-1}\boldsymbol{XAMX}^{T}\rightarrow\boldsymbol{\Omega}_{2},\ \ \ \ %
p^{-1}\boldsymbol{XMA}^{2}\boldsymbol{MX}^{T}\rightarrow\boldsymbol{\Omega
}_{3}, \label{conditionGLS}%
\end{equation}
where only the first and third conditions above are assumed for Model I and only the first and the second are assumed
for Model II. Then we have%
\begin{equation}
\sup\limits_{\boldsymbol{B}\in\mathcal{B}}\left\vert \mathrm{URE}%
_{\boldsymbol{M}}\left(  \boldsymbol{B}\right)  -l_{p}\left( \boldsymbol{\theta
},\boldsymbol{\hat{\theta}}^{\boldsymbol{B},\boldsymbol{\hat{\mu}%
}^{\boldsymbol{M}}} \right) \right\vert \rightarrow0\text{ in }L^{1}\text{ as
}p\rightarrow\infty. \label{unifconvGLS}%
\end{equation}
As a corollary, for any estimator $\boldsymbol{\hat{\theta}}^{\boldsymbol{\hat
{B}}_{p},\boldsymbol{\hat{\mu}}^{\boldsymbol{M}}}=\boldsymbol{\hat{B}}%
_{p}\left( \boldsymbol{A}+\boldsymbol{\hat{B}}_{p} \right) ^{-1}\boldsymbol{Y}%
+\boldsymbol{A}\left( \boldsymbol{A}+\boldsymbol{\hat{B}}_{p} \right) ^{-1}\boldsymbol{\hat
{\mu}}^{\boldsymbol{M}}$ with $\boldsymbol{\hat{B}}_{p}\in\mathcal{B}$, we
always have%
\begin{align*}
\lim\limits_{p\rightarrow\infty}\mathbb{P}\left(  l_{p}\left(
\boldsymbol{\theta},\boldsymbol{\hat{\theta}}_{\boldsymbol{M}}^{\mathrm{URE}%
}\right)  \geq l_{p}\left(  \boldsymbol{\theta},\boldsymbol{\hat{\theta}%
}^{\boldsymbol{\hat{B}}_{p},\boldsymbol{\hat{\mu}}^{\boldsymbol{M}}}\right)
+\epsilon\right)   &  =0\ \ \ \ \forall\epsilon>0,\\
\limsup\limits_{p\rightarrow\infty}\left(  R_{p}\left(  \boldsymbol{\theta
},\boldsymbol{\hat{\theta}}_{\boldsymbol{M}}^{\mathrm{URE}}\right)
-R_{p}\left(  \boldsymbol{\theta},\boldsymbol{\hat{\theta}}^{\boldsymbol{\hat
{B}}_{p},\boldsymbol{\hat{\mu}}^{\boldsymbol{M}}}\right)  \right)   &  \leq0.
\end{align*}
\end{theorem}

\begin{remark}
For shrinking toward $\boldsymbol{\hat{\mu}}^{\mathrm{OLS}}$,
where $\boldsymbol{M}=\boldsymbol{I}_{p}$, we know from
Lemma \ref{Lma.: WLS is in the restriction class} that $\boldsymbol{\hat{\mu}%
}^{\mathrm{OLS}}$ is automatically in $\mathcal{L}$, so we only need one more
condition $p^{-1}\boldsymbol{XA}^{2}\boldsymbol{X}^{T}\rightarrow
\boldsymbol{\Omega}_{3}$ for Model I. For shrinking toward $\boldsymbol{\hat{\mu}%
}^{\mathrm{WLS}}$, where
$\boldsymbol{M}=\boldsymbol{A}^{-1}$, (\ref{conditionGLS}) is the same as
the conditions (E) and (F) of Theorem \ref{Thm.: Main Theorem}, so additionally we
only need to assume $(\mathrm{A}^{\prime})$ of Lemma
\ref{Lma.: WLS is in the restriction class} and (F) for Model I.
\end{remark}

\section{Semiparametric URE Estimators}

\label{Sec.: Semiparametric URE}

We have established the (asymptotic) optimality of the URE estimators
$\boldsymbol{\hat{\theta}}^{\mathrm{URE}}$ and $\boldsymbol{\hat{\theta}%
}_{\boldsymbol{M}}^{\mathrm{URE}}$ in the previous section. One limitation of
the result is that the class over which the URE estimators are optimal is
specified by a parametric form: $\boldsymbol{B}=\lambda\boldsymbol{C}$
($0\leq\lambda\leq\infty$) in equation (\ref{Generic Shrinkage Estimator}),
where $\boldsymbol{C}=\boldsymbol{I}_{p}$ for Model I and $\boldsymbol{C}%
=\boldsymbol{X}^{T}\boldsymbol{WX}$ for Model II. Aiming to provide a more
flexible and, at the same time, efficient estimation procedure, we consider in
this section a class of semiparametric shrinkage estimators. Our consideration
is inspired by \cite{Xie&Kou&Brown-2012}.

\subsection{Semiparametric URE Estimator under Model I}

To motivate the semiparametric shrinkage estimators, let us first revisit the
Bayes estimator $\boldsymbol{\hat{\theta}}^{\lambda,\boldsymbol{\mu}}$ under
Model I, as given in (\ref{Model I Shrinkage estimator}). It is seen that the
Bayes estimate of each mean parameter $\theta_{i}$ is obtained by shrinking
$Y_{i}$ toward the linear estimate $\mu_{i}=\boldsymbol{X}_{i}^{T}%
\boldsymbol{\beta}$, and that the amount of shrinkage is governed by $A_{i}$,
the variance: the larger the variance, the stronger is the shrinkage. This
feature makes intuitive sense.

With this observation in mind, we consider the following shrinkage estimators
under Model I:%
\[
\hat{\theta}_{i}^{\boldsymbol{b},\boldsymbol{\mu}}=\left(  1-b_{i}\right)
Y_{i}+b_{i}\mu_{i},\ \ \ \ \text{with }\boldsymbol{\mu}\in\mathcal{L}%
_{\mathrm{row}}\left(  \boldsymbol{X}\right)  ,
\]
where $\boldsymbol{b}$ satisfies the monotonic constraint%
\[
\mathrm{MON}\left(  \boldsymbol{A}\right)  :b_{i}\in\left[  0,1\right]
,\ b_{i}\leq b_{j}\text{ whenever }A_{i}\leq A_{j}.
\]
$\mathrm{MON}\left(  \boldsymbol{A}\right)  $ asks the estimator to shrink
more for an observation with a larger variance. Since other than this
intuitive requirement, we do not post any parametric restriction on $b_{i}$,
this class of estimators is semiparametric in nature.

Following the optimality result for the parametric case, we want to
investigate, for such a general estimator $\boldsymbol{\hat{\theta}%
}^{\boldsymbol{b},\boldsymbol{\mu}}$ with $\boldsymbol{b}\in\mathrm{MON}%
\left(  \boldsymbol{A}\right)  $ and $\boldsymbol{\mu}\in\mathcal{L}%
_{\mathrm{row}}\left(  \boldsymbol{X}\right)  $, whether there exists an
optimal choice of $\boldsymbol{b}$ and $\boldsymbol{\mu}$. In fact, we will
see shortly that such an optimal choice exists, and this asymptotically
optimal choice is again characterized by an unbiased risk estimate (URE). For
a general estimator $\boldsymbol{\hat{\theta}}^{\boldsymbol{b},\boldsymbol{\mu
}}$ with \emph{fixed} $\boldsymbol{b}$ and $\boldsymbol{\mu}\in\mathcal{L}%
_{\mathrm{row}}\left(  \boldsymbol{X}\right)  $, an unbiased estimate of its
risk $R_{p}(\boldsymbol{\theta},\boldsymbol{\hat{\theta}}^{\boldsymbol{b}%
,\boldsymbol{\mu}})$ is
\[
\mathrm{URE}^{SP}\left(  \boldsymbol{b},\boldsymbol{\mu}\right)  =\dfrac{1}%
{p}\left\Vert \mathrm{diag}\left(  \boldsymbol{b}\right)  \left(
\boldsymbol{Y}-\boldsymbol{\mu}\right)  \right\Vert ^{2}+\dfrac{1}%
{p}\mathrm{tr}\left(  \boldsymbol{A}-2\mathrm{diag}\left(  \boldsymbol{b}%
\right)  \boldsymbol{A}\right)  ,
\]
which can be easily seen by taking $\boldsymbol{B=A}(\mathrm{diag}\left(
\boldsymbol{b}\right)  ^{-1}-\boldsymbol{I}_{p})$ in (\ref{Generic URE}). Note
that we use the superscript \textquotedblleft\textit{SP}\textquotedblright%
\ (semiparametric) to denote it. Minimizing over $\boldsymbol{b}$ and
$\boldsymbol{\mu}$ leads to the semiparametric URE estimator $\boldsymbol{\hat
{\theta}}_{SP}^{\mathrm{URE}}$, defined by%
\begin{equation}
\boldsymbol{\hat{\theta}}_{SP}^{\mathrm{URE}}=(\boldsymbol{I}_{p}%
-\mathrm{diag}(\boldsymbol{\hat{b}}_{SP}^{\mathrm{URE}}))\boldsymbol{Y}%
+\mathrm{diag}(\boldsymbol{\hat{b}}_{SP}^{\mathrm{URE}})\boldsymbol{\hat{\mu}%
}_{SP}^{\mathrm{URE}}\mathbf{,} \label{SPURE}%
\end{equation}
where%
\[
\left(  \boldsymbol{\hat{b}}_{SP}^{\mathrm{URE}},\boldsymbol{\hat{\mu}}%
_{SP}^{\mathrm{URE}}\right)  =\operatorname*{argmin}\limits_{\boldsymbol{b}%
\in\mathrm{MON}\left(  \boldsymbol{A}\right)  ,\ \boldsymbol{\mu}%
\in\mathcal{L}_{\mathrm{row}}\left(  \boldsymbol{X}\right)  }\mathrm{URE}%
^{SP}\left(  \boldsymbol{b},\boldsymbol{\mu}\right)  .
\]

\begin{theorem}
Assume conditions (A)-(E). Then under Model I we have%
\[
\sup\limits_{\boldsymbol{b}\in\mathrm{MON}\left(  \boldsymbol{A}\right)
,\;\boldsymbol{\mu}\in\mathcal{L}}\left\vert \mathrm{URE}^{SP}\left(
\boldsymbol{b},\boldsymbol{\mu}\right)  -l_{p}\left(  \boldsymbol{\theta
},\boldsymbol{\hat{\theta}}^{\boldsymbol{b},\boldsymbol{\mu}}\right)
\right\vert \rightarrow0\text{ in }L^{1}\text{ as }p\rightarrow\infty.
\]
As a corollary, for any estimator $\boldsymbol{\hat{\theta}}^{\boldsymbol{\hat
{b}_{p}},\boldsymbol{\hat{\mu}}_{p}}= ( \boldsymbol{I}_{p}-\mathrm{diag}%
( \boldsymbol{\hat{b}}_{p} ) ) \boldsymbol{Y}+\mathrm{diag} ( \boldsymbol{\hat{b}%
}_{p} ) \boldsymbol{\hat{\mu}}_{p}$ with $\boldsymbol{\hat{b}}_{p}\in\mathrm{MON}\left(
\boldsymbol{A}\right)  $ and $\boldsymbol{\hat{\mu}}_{p}\in\mathcal{L}$, we always have%
\begin{align*}
\lim\limits_{p\rightarrow\infty}\mathbb{P}\left(  l_{p}\left(
\boldsymbol{\theta},\boldsymbol{\hat{\theta}}_{SP}^{\mathrm{URE}}\right)  \geq
l_{p}\left(  \boldsymbol{\theta},\boldsymbol{\hat{\theta}}^{\boldsymbol{\hat
{b}}_{p},\boldsymbol{\hat{\mu}}_{p}}\right)  +\epsilon\right)   &  =0\ \ \ \ \forall
\epsilon>0,\\
\limsup\limits_{p\rightarrow\infty}\left(  R_{p}\left( \boldsymbol{\theta
},\boldsymbol{\hat{\theta}}_{SP}^{\mathrm{URE}} \right) -R_{p}\left( \boldsymbol{\theta
},\boldsymbol{\hat{\theta}}^{\boldsymbol{\hat{b}}_{p},\boldsymbol{\hat{\mu}}_{p}%
} \right) \right)   &  \leq0.
\end{align*}
\end{theorem}

The proof is the same as the proofs of Theorem \ref{Thm.: Main Theorem} and
Corollary \ref{Cor.: Risk properties} for the case of Model I except that we
replace each term of $A_{i}/(\lambda+A_{i})$ by $b_{i}$.

\subsection{Semiparametric URE Estimator Under Model II}

We saw in Sec.~\ref{Sec.: Heteroscedastic Hierarchical Linear Models} that,
under Model II, shrinkage\ is achieved by shrinking the regression coefficient
from the weighted least squares estimate $\boldsymbol{\hat{\beta}%
}^{\mathrm{WLS}}$ toward the prior coefficient $\boldsymbol{\beta}_{0}$. This
suggests us to formulate the semiparametric estimators through the regression
coefficient. The Bayes estimate of the regression coefficient is%
\[
\boldsymbol{\hat{\beta}}^{\lambda,\boldsymbol{\beta}_{0}}=\lambda
\boldsymbol{W}(\lambda\boldsymbol{W}+\boldsymbol{V})^{-1}\boldsymbol{\hat
{\beta}}^{\mathrm{WLS}}+\boldsymbol{V}\left(  \lambda\boldsymbol{W}%
+\boldsymbol{V}\right)  ^{-1}\boldsymbol{\beta}_{0},\ \ \ \ \text{with
}\boldsymbol{V}=(\boldsymbol{XA}^{-1}\boldsymbol{X}^{T})^{-1}%
\]
as shown in (\ref{Model II Shrinkage estimator}). Applying the spectral
decomposition on $\boldsymbol{W}^{-1/2}\boldsymbol{VW}^{-1/2}$ gives
$\boldsymbol{W}^{-1/2}\boldsymbol{VW}^{-1/2}=\boldsymbol{U\Lambda
}\boldsymbol{U}^{T}$, where $\boldsymbol{\Lambda}=\mathrm{diag}\left(
d_{1},...,d_{k}\right)  $ with $d_{1}\leq\cdots\leq d_{k}$. Using this
decomposition, we can rewrite the regression coefficient as%
\[
\boldsymbol{\hat{\beta}}^{\lambda,\boldsymbol{\beta}_{0}}=\lambda
\boldsymbol{W}^{1/2}\boldsymbol{U}\left(  \lambda\boldsymbol{I}_{k}%
+\boldsymbol{\Lambda}\right)  ^{-1}\boldsymbol{U}^{T}\boldsymbol{W}%
^{-1/2}\boldsymbol{\hat{\beta}}^{\mathrm{WLS}}+\boldsymbol{W}^{1/2}%
\boldsymbol{U\Lambda}\left(  \lambda\boldsymbol{I}_{k}+\boldsymbol{\Lambda
}\right)  ^{-1}\boldsymbol{U}^{T}\boldsymbol{W}^{-1/2}\boldsymbol{\beta}_{0}.
\]
If we denote $\boldsymbol{Z}=\boldsymbol{U}^{T}\boldsymbol{W}^{1/2}%
\boldsymbol{X}$ as the transformed covariate matrix, the estimate
$\boldsymbol{\hat{\theta}}^{\lambda,\boldsymbol{\beta}_{0}}=\boldsymbol{X}%
^{T}\boldsymbol{\hat{\beta}}^{\lambda,\boldsymbol{\beta}_{0}}$ of
$\boldsymbol{\theta}$ can be rewritten as%
\[
\boldsymbol{\hat{\theta}}^{\lambda,\boldsymbol{\beta}_{0}}=\boldsymbol{Z}%
^{T}\left(  \lambda\left(  \lambda\boldsymbol{I}_{k}+\boldsymbol{\Lambda
}\right)  ^{-1}\boldsymbol{U}^{T}\boldsymbol{W}^{-1/2}\boldsymbol{\hat{\beta}%
}^{\mathrm{WLS}}+\boldsymbol{\Lambda}\left(  \lambda\boldsymbol{I}%
_{k}+\boldsymbol{\Lambda}\right)  ^{-1}\boldsymbol{U}^{T}\boldsymbol{W}%
^{-1/2}\boldsymbol{\beta}_{0}\right)  .
\]
Now we see that $\lambda\left(  \lambda\boldsymbol{I}_{k}+\boldsymbol{\Lambda
}\right)  ^{-1}=\mathrm{diag}(\lambda/\left(  \lambda+d_{i}\right)  )$ plays
the role as the shrinkage factor. The larger the value of $d_{i}$, the smaller
$\lambda/\left(  \lambda+d_{i}\right)  $, i.e., the stronger the shrinkage
toward $\boldsymbol{\beta}_{0}$. Thus, $d_{i}$ can be viewed as the effective
\textquotedblleft variance\textquotedblright\ component for the $i$-th
regression coefficient (under the transformation). This observation motivates
us to consider semiparametric shrinkage estimators of the following form%
\begin{align}
\boldsymbol{\hat{\theta}}^{\boldsymbol{b},\boldsymbol{\beta}_{0}}  &
=\boldsymbol{Z}^{T}\left(  \left(  \boldsymbol{I}_{k}-\mathrm{diag}\left(
\boldsymbol{b}\right)  \right)  \boldsymbol{U}^{T}\boldsymbol{W}%
^{-1/2}\boldsymbol{\hat{\beta}}^{\mathrm{WLS}}+\mathrm{diag}\left(
\boldsymbol{b}\right)  \boldsymbol{U}^{T}\boldsymbol{W}^{-1/2}%
\boldsymbol{\beta}_{0}\right) \nonumber\\
&  =\boldsymbol{Z}^{T}\left(  \left(  \boldsymbol{I}_{k}-\mathrm{diag}\left(
\boldsymbol{b}\right)  \right)  \boldsymbol{\Lambda ZA}^{-1}\boldsymbol{Y}%
+\mathrm{diag}\left(  \boldsymbol{b}\right)  \boldsymbol{U}^{T}\boldsymbol{W}%
^{-1/2}\boldsymbol{\beta}_{0}\right)  , \label{SPUREModelII}%
\end{align}
where $\boldsymbol{b}$ satisfies the following monotonic constraint%
\[
\mathrm{MON}\left(  \boldsymbol{D}\right)  :b_{i}\in\left[  0,1\right]
,\ b_{i}\leq b_{j}\text{ whenever }d_{i}\leq d_{j}.
\]
This constraint captures the intuition that, the larger the effective
variance, the stronger is the shrinkage.

For \emph{fixed} $\boldsymbol{b}$ and $\boldsymbol{\beta}_{0}$, an unbiased
estimate of the risk $R_{p}(\boldsymbol{\theta},\boldsymbol{\hat{\theta}%
}^{\boldsymbol{b},\boldsymbol{\beta}_{0}})$ is%
\begin{align*}
\mathrm{URE}^{SP}\left(  \boldsymbol{b},\boldsymbol{\beta}_{0}\right)   &
=\dfrac{1}{p}\left\Vert \boldsymbol{Z}^{T}\left(  \boldsymbol{I}%
_{k}-\mathrm{diag}\left(  \boldsymbol{b}\right)  \right)  \boldsymbol{\Lambda
ZA}^{-1}\boldsymbol{Y}+\boldsymbol{Z}^{T}\mathrm{diag}\left(  \boldsymbol{b}%
\right)  \boldsymbol{U}^{T}\boldsymbol{W}^{-1/2}\boldsymbol{\beta}%
_{0}-\boldsymbol{Y}\right\Vert ^{2}\\
&  +\dfrac{1}{p}\mathrm{tr}\left(  2\boldsymbol{Z}^{T}\left(  \boldsymbol{I}%
_{k}-\mathrm{diag}\left(  \boldsymbol{b}\right)  \right)  \boldsymbol{\Lambda
Z}-\boldsymbol{A}\right)  ,
\end{align*}
which can be shown using the bias-variance decomposition of the mean squared
error. Minimizing it gives the URE estimate of $\left(  \boldsymbol{b}%
,\boldsymbol{\beta}_{0}\right)  $:%
\[
\left(  \boldsymbol{\hat{b}}_{SP}^{\mathrm{URE}},\left(  \boldsymbol{\hat
{\beta}}_{0}\right)  _{SP}^{\mathrm{URE}}\right)  =\operatorname*{argmin}%
\limits_{\boldsymbol{b}\in\mathrm{MON}\left(  \boldsymbol{D}\right)
,\ \boldsymbol{\beta}_{0}\in\mathbb{R}^{k}}\mathrm{URE}^{SP}\left(
\boldsymbol{b},\boldsymbol{\beta}_{0}\right)  ,
\]
which upon plugging into (\ref{SPUREModelII}) yields the semiparametric URE
estimator $\boldsymbol{\hat{\theta}}_{SP}^{\mathrm{URE}}$ under Model II.

\begin{theorem}
Assume conditions (A), (D)-(G). Then under Model II we have%
\[
\sup\limits_{\boldsymbol{b}\in\mathrm{MON}\left(  \boldsymbol{D}\right)
,\;\boldsymbol{X}^{T}\boldsymbol{\beta}_{0}\in\mathcal{L}}\left\vert
\mathrm{URE}^{SP}\left(  \boldsymbol{b},\boldsymbol{\beta}_{0}\right)
-l_{p}\left(  \boldsymbol{\theta},\boldsymbol{\hat{\theta}}^{\boldsymbol{b}%
,\boldsymbol{\beta}_{0}}\right)  \right\vert \rightarrow0\text{ in }%
L^{1}\text{ as }p\rightarrow\infty.
\]
As a corollary, for any estimator $\boldsymbol{\hat{\theta}}^{\boldsymbol{\hat
{b}_{p}},\boldsymbol{\hat{\beta}}_{0,p}}$ obtained from (\ref{SPUREModelII}) with
$\boldsymbol{\hat{b}}_{p}\in\mathrm{MON}\left(  \boldsymbol{D}\right)
$ and $\boldsymbol{X}^{T}\boldsymbol{\hat{\beta}}_{0}\in\mathcal{L}$, we always have%
\begin{align*}
\lim\limits_{p\rightarrow\infty}\mathbb{P}\left(  l_{p}\left(
\boldsymbol{\theta},\boldsymbol{\hat{\theta}}_{SP}^{\mathrm{URE}}\right)  \geq
l_{p}\left(  \boldsymbol{\theta},\boldsymbol{\hat{\theta}}^{\boldsymbol{\hat
{b}}_{p},\boldsymbol{\hat{\beta}}_{0,p}}\right)  +\epsilon\right)   &
=0\ \ \ \ \forall\epsilon>0,\\
\limsup\limits_{p\rightarrow\infty}\left(  R_{p}\left( \boldsymbol{\theta
},\boldsymbol{\hat{\theta}}_{SP}^{\mathrm{URE}} \right) -R_{p}\left( \boldsymbol{\theta
},\boldsymbol{\hat{\theta}}^{\boldsymbol{\hat{b}}_{p},\boldsymbol{\hat{\beta}}%
_{0,p}} \right) \right)   &  \leq0.
\end{align*}
\end{theorem}

The proof of the theorem is essentially identical to those of Theorem
\ref{Thm.: Main Theorem} and Corollary \ref{Cor.: Risk properties} for the
case of Model II except that we replace each $d_{i}/(\lambda+d_{i})$ by
$b_{i}$.

\section{Simulation Study}

\label{Sec.: Simulation}

In this section, we conduct simulations to study the performance of the URE
estimators. For the sake of space, we will focus on Model I. The four URE
estimators are the parametric $\boldsymbol{\hat{\theta}}^{\mathrm{URE}}$ of
equation (\ref{genericUREest}), the parametric $\boldsymbol{\hat{\theta}%
}_{\boldsymbol{M}}^{\mathrm{URE}}$ of equation (\ref{UREMEst}) that shrinks
toward the OLS estimate $\boldsymbol{\hat{\mu}}^{\mathrm{OLS}}$ (i.e., the
matrix $\boldsymbol{M}=\boldsymbol{I}_{p}$), the semiparametric
$\boldsymbol{\hat{\theta}}_{SP}^{\mathrm{URE}}$ of equation (\ref{SPURE}), and
the semiparametric $\boldsymbol{\hat{\theta}}_{SP}^{\mathrm{URE}%
,\,\mathrm{OLS}}$ that shrinks toward $\boldsymbol{\hat{\mu}}^{\mathrm{OLS}}$,
which is formed similarly to $\boldsymbol{\hat{\theta}}_{\boldsymbol{M}%
}^{\mathrm{URE}}$ by replacing $A_{i}/(\lambda+A_{i})$ with a sequence
$\boldsymbol{b}\in\mathrm{MON}\left(  \boldsymbol{A}\right)  $. The
competitors here are the two empirical Bayes estimators $\boldsymbol{\hat
{\theta}}^{\mathrm{EBMLE}}$ and $\boldsymbol{\hat{\theta}}^{\mathrm{EBMOM}}$,
and the positive part James-Stein estimator $\boldsymbol{\hat{\theta}%
}^{\mathrm{JS}+}$ as described in \cite{Brown-2008, Morris&Lysy-2012}:%
\[
\hat{\theta}_{i}^{\mathrm{JS+}}=\hat{\mu}_{i}^{\mathrm{WLS}}+\left(
1-\frac{p-k-2}{\sum_{i=1}^{p}\left(  Y_{i}-\hat{\mu}_{i}^{\mathrm{WLS}%
}\right)  ^{2}/A_{i}}\right)  ^{+}\left(  Y_{i}-\hat{\mu}_{i}^{\mathrm{WLS}}
\right)  .
\]

As a reference, we also compare these shrinkage estimators with
$\boldsymbol{\tilde{\theta}}^{\mathrm{OR}}$, the parametric oracle risk (OR)
estimator, defined as plugging $\tilde{\lambda}^{\mathrm{OR}}\boldsymbol{I}%
_{p}$ and $\boldsymbol{\tilde{\mu}}^{\mathrm{OR}}$ into equation
(\ref{Generic Shrinkage Estimator}), where%
\[
\left(  \tilde{\lambda}^{\mathrm{OR}},\boldsymbol{\tilde{\mu}}^{\mathrm{OR}%
}\right)  =\operatorname*{argmin}\limits_{0\leq\lambda\leq\infty
,\;\boldsymbol{\mu}\in\mathcal{L}_{\mathrm{row}}\left(  \boldsymbol{X}\right)
}R_{p}\left(  \boldsymbol{\theta},\boldsymbol{\hat{\theta}}^{\lambda
,\boldsymbol{\mu}}\right)
\]
and the expression of $R_{p}(\boldsymbol{\theta},\boldsymbol{\hat{\theta}%
}^{\lambda,\boldsymbol{\mu}})$ is given in (\ref{Generic Risk}) with
$\boldsymbol{B}=\lambda\boldsymbol{I}_{p}$. The oracle risk estimator
$\boldsymbol{\tilde{\theta}}^{\mathrm{OR}}$ cannot be used without the
knowledge of $\boldsymbol{\theta}$, but it does provide a sensible lower bound
of the risk achievable by any shrinkage estimator with the given parametric form.

\begin{figure}[t]
\centering
\includegraphics[width=1\linewidth]{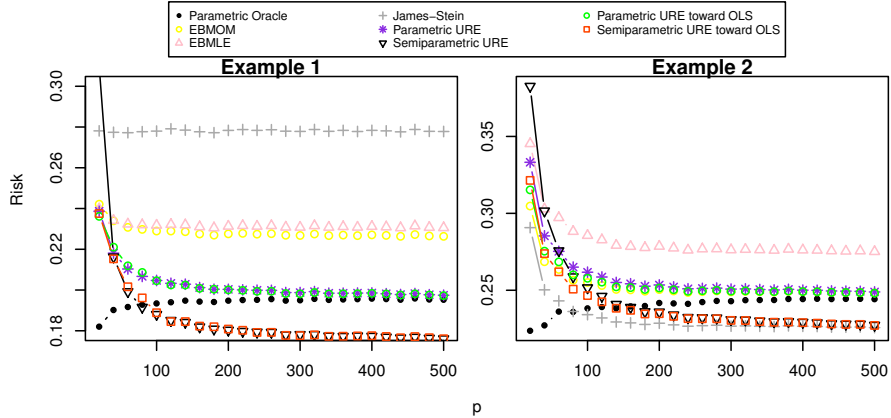}\caption{Comparison
of the risks of different shrinkage estimators for the two simulation
examples.}%
\label{fig.: Risk Performance Simulation}%
\end{figure}

For each simulation, we draw $\left(  A_{i},\theta_{i}\right)  $
($i=1,2,...,p$) independently from a distribution $\pi\left(  A_{i},\theta
_{i}|\boldsymbol{X}_{i},\boldsymbol{\beta}\right)  $ and then draw $Y_{i}$
given $\left(  A_{i},\theta_{i}\right)  $. The shrinkage estimators are then
applied to the generated data. This process is repeated $5000$ times. The
sample size $p$ is chosen to vary from $20$ to $500$ with an increment of
length $20$. In the simulation, we fix a true but unknown $\boldsymbol{\beta
}=\left(  -1.5,4,-3\right)  ^{T}$ and a known covariates $\boldsymbol{X}$,
whose each element is randomly generated from $\mathrm{Unif}\left(
-10,10\right)  $. The risk performance of the different shrinkage estimators
is given in Figure \ref{fig.: Risk Performance Simulation}.

\textit{\textbf{{Example 1.}}} The setting in this example is chosen in such a
way that it reflects grouping in the data:%
\begin{align*}
A_{i}  &  \sim0.5\cdot1_{\left\{  A_{i}=0.1\right\}  }+0.5\cdot1_{\left\{
A_{i}=0.5\right\}  };\ \\
\theta_{i}|A_{i}  &  \sim N\left(  2\cdot1_{\left\{  A_{i}=0.1\right\}
}+\boldsymbol{X}_{i}^{T}\boldsymbol{\beta},0.5^{2}\right)  ;\ Y_{i}\sim
N\left(  \theta_{i},A_{i}\right)  .
\end{align*}
Here the normality for the sampling distribution of $Y_{i}$'s is asserted. We
can see that the four URE estimators perform much better than the two
empirical Bayes ones and the James-Stein estimator. Also notice that both of
the two (parametric and semiparametric) URE estimators that shrink towards
$\boldsymbol{\hat{\mu}}^{\mathrm{OLS}}$ is almost as good as the other two
with general data-driven shrinkage location---largely due to the existence of
covariate information. We note that this is quite different from the case of
\cite{Xie&Kou&Brown-2012}, where without the covariate information the
estimator that shrinks toward the grand mean of the data performs
significantly worse than the URE estimator with general data-driven shrinkage location.

\textit{\textbf{{Example 2.}}} In this example, we allow $Y_{i}$ to depart
from the normal distribution to illustrate that the performance of those URE
estimators does not rely on the normality assumption:%
\begin{align*}
A_{i}  &  \sim\mathrm{Unif}\left(  0.1,1\right)  ;\ \theta_{i}=A_{i}%
+\boldsymbol{X}_{i}^{T}\boldsymbol{\beta};\\
Y_{i}  &  \sim\mathrm{Unif}(\theta_{i}-\sqrt{3}A_{i},\theta_{i}+\sqrt{3}%
A_{i}).
\end{align*}
As expected, the four URE estimators perform better or at least as good as the
empirical Bayes estimators. The EBMLE estimator performs the worst due to its
sensitivity on the normality assumption. We notice that the EBMOM estimator in
this example has comparable performance with the two parametric URE
estimators, which makes sense as moment estimates are more robust to the
sampling distribution. An interesting feature that we find in this example is
that the positive part James-Stein estimator can beat the parametric oracle
risk estimator and perform better than all the other shrinkage estimators for
small or moderate $p$, even though the semiparametric URE estimators will
eventually surpass the James-Stein estimator, as dictated by the asymptotic
theory for large $p$. This feature of the James-Stein estimate is again quite
different from the non-regression setting discussed in
\cite{Xie&Kou&Brown-2012}, where the James-Stein estimate performs the worst
throughout all of their examples. In both of our examples only the
semiparametric URE estimators are robust to the different levels of
heteroscedasticity. 

We can conclude from these two simulation examples that the semiparametric URE
estimators give competitive performance and are robust to the misspecification
of the sampling distribution and the different levels of the
heteroscedasticity. They thus could be useful tools in analyzing large-scale
data for applied researchers.

\section{Empirical Analysis}

\label{Sec.: Baseball data analysis}

In this section, we study the baseball data set of \cite{Brown-2008}. This
data set consists of the batting records for all the Major League Baseball
players in the 2005 season. As in \cite{Brown-2008} and
\cite{Xie&Kou&Brown-2012}, we build a given shrinkage estimator based on the
data in the first half season and use it to predict the second half season,
which can then be checked against the true record of the second half season.
For each player, let the number of at-bats be $N$ and the successful number of
batting be $H$, then we have $H_{ij}\sim Binomial(N_{ij},p_{j})$, where
$i=1,2$ is the season indicator and $j=1,\cdots,p$ is the player indicator. We
use the following variance-stabilizing transformation \cite{Brown-2008} before
applying the shrinkage estimators
\[
Y_{ij}=\arcsin\sqrt{\frac{H_{ij}+1/4}{N_{ij}+1/2}},
\]
which gives $Y_{ij}\dot{\sim}N(\theta_{j},(4N_{ij})^{-1})$, $\theta
_{j}=\arcsin\sqrt{p_{j}}$. We use%
\[
\mathrm{TSE}(\boldsymbol{\hat{\theta}})=\sum_{j}(Y_{2j}-\hat{\theta}_{j}%
)^{2}-\sum\limits_{j}\frac{1}{4N_{2j}}.
\]
as the error measurement for the prediction \cite{Brown-2008}.

\subsection{Shrinkage Estimation with Covariates}

As indicated in \cite{Xie&Kou&Brown-2012}, there exists a significant positive
correlation between the player's batting ability and his total number of
at-bats. Intuitively, a better player will be called for batting more
frequently; thus, the total number of at-bats will serve as the main covariate
in our analysis. The other covariate in the data set is the categorical
variable of a player being a pitcher or not.

Table \ref{tab.: Baseball data prediction} summarizes the result, where the
shrinkage estimators are applied three times---to all the players, the
pitchers only, and the non-pitchers only. We use all the covariate information
(number of at-bats in the first half season and being a pitcher or not) in the
first analysis, whereas in the second and the third analyses we only use the
number of at-bats as the covariate. The values reported are ratios of the
error of a given estimator to that of the benchmark naive estimator, which
simply uses the first half season $Y_{1j}$ to predict the second half $Y_{2j}%
$. Note that in Table \ref{tab.: Baseball data prediction}, if no covariate is
involved (i.e., when $\boldsymbol{X}=\left[  1|\cdots|1\right]  $), the
$\mathrm{OLS}$ reduces to the grand mean of the training data as in
\cite{Xie&Kou&Brown-2012}.

\begin{table}[ptb]
\centering$%
\begin{tabular}
[c]{rrrrrrr}
& \multicolumn{2}{c}{All} & \multicolumn{2}{c}{Pichers} &
\multicolumn{2}{c}{Non-pichers}\\\hline\hline
\multicolumn{1}{l}{$p$ for estimation} & \multicolumn{2}{c}{567} &
\multicolumn{2}{c}{81} & \multicolumn{2}{c}{486}\\
\multicolumn{1}{l}{$p$ for validation} & \multicolumn{2}{c}{499} &
\multicolumn{2}{c}{64} & \multicolumn{2}{c}{435}\\\hline
\multicolumn{1}{l}{Covariates?} & No & Yes & No & Yes & No & Yes\\
\multicolumn{1}{l}{Naive} & 1 & $\mathrm{NA}$ & 1 & $\mathrm{NA}$ & 1 &
$\mathrm{NA}$\\
\multicolumn{1}{l}{Ordinary least squares ($\mathrm{OLS}$)} & 0.852 & 0.242 &
0.127 & 0.115 & 0.378 & 0.333\\
\multicolumn{1}{l}{Weighted least squares ($\mathrm{WLS}$)} & 1.074 & 0.219 &
0.127 & 0.087 & 0.468 & 0.290\\
\multicolumn{1}{l}{Parametric $\mathrm{EBMOM}$} & 0.593 & 0.194 & 0.129 &
0.117 & 0.387 & \textbf{{0.256}}\\
\multicolumn{1}{l}{Parametric $\mathrm{EBMLE}$} & 0.902 & 0.207 & 0.117 &
0.096 & 0.398 & 0.277\\
\multicolumn{1}{l}{James-Stein} & 0.525 & \textbf{{0.184}} & 0.164 & 0.142 &
0.359 & 0.262\\
\multicolumn{1}{l}{Parametric $\mathrm{URE}$ toward $\mathrm{OLS}$} & 0.505 &
0.203 & 0.123 & 0.124 & 0.278 & 0.300\\
\multicolumn{1}{l}{Parametric $\mathrm{URE}$ toward $\mathrm{WLS}$} & 0.629 &
0.188 & 0.127 & 0.112 & 0.385 & 0.268\\
\multicolumn{1}{l}{Parametric $\mathrm{URE}$} & 0.422 & 0.215 & 0.123 &
0.130 & 0.282 & 0.310\\
\multicolumn{1}{l}{Semiparametric $\mathrm{URE}$ toward $\mathrm{OLS}$} &
0.409 & 0.197 & 0.081 & 0.097 & 0.261 & 0.299\\
\multicolumn{1}{l}{Semiparametric $\mathrm{URE}$ toward $\mathrm{WLS}$} &
0.499 & \textbf{{0.184}} & 0.098 & \textbf{{0.083}} & 0.336 & \textbf{{0.256}%
}\\
\multicolumn{1}{l}{Semiparametric $\mathrm{URE}$} & 0.419 & 0.201 & 0.077 &
0.126 & 0.278 & 0.314
\end{tabular}
\ \ $\caption{Prediction errors of batting averages using different shrinkage
estimators. Bold numbers highlight the best performance with covariate(s) in
each case.}%
\label{tab.: Baseball data prediction}%
\end{table}

\subsection{Discussion of the numerical result}

There are several interesting observations from Table
\ref{tab.: Baseball data prediction}.

(i) A quick glimpse shows that including the covariate information improves
the performance of essentially all shrinkage estimators. This suggests that in
practice incorporating good covariates would significantly improve the
estimation and prediction.

(ii) In general, shrinking towards WLS provides much better performance than
shrinking toward OLS or a general data-driven location. This indicates the
importance of a good choice of the shrinkage location in a practical problem.
An improperly chosen shrinkage location might even negatively impact the
performance. The reason that shrinking towards a general data-driven location
is not as good as shrinking toward WLS\ is probably due to that the sample
size is not large enough for the asymptotics to take effect.

(iii) Table \ref{tab.: Baseball data prediction} also shows the advantage of
semiparametric URE estimates. For each fixed shrinkage location type (toward
$\mathrm{OLS}$, $\mathrm{WLS}$, or general), the semiparametric URE estimator
performs almost always better than their parametric counterparts. The only one
exception is in the non-pitchers only case with the general data-driven
location, but even there the performance difference is ignorable.

(iv) The best performance in all three cases (all the players, the pitchers
only, and the non-pitchers only) comes from the semiparametric URE estimator
that shrinks toward WLS.

(v) The James-Stein estimator \emph{with} covariates performs quite well
except in the pitchers only case, which is in sharp contrast with the
performance of the James-Stein estimator \emph{without} covariates. This again
highlights the importance of covariate information. In the pitchers only case,
the James-Stein performs the worst no matter one includes the covariates or
not. This can be attributed to the fact that the covariate information (the
total number of at-bats) is very weak for the pitchers only case; in the case
of weak covariate information, how to properly estimate the shrinkage factors
becomes the dominating issue, and the fact that the James-Stein estimator has
only \emph{one} uniform shrinkage factor makes it not competitive.

\subsection{Shrinkage Factors}

\begin{figure}[t]
\centering\includegraphics[width=1\linewidth]{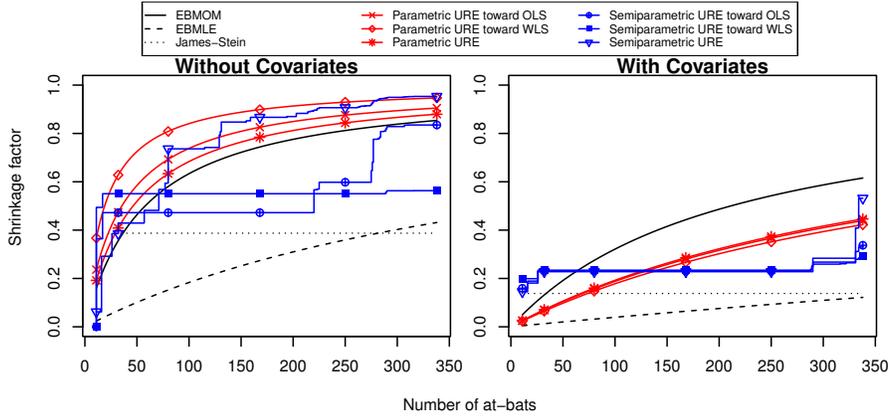}\caption{Plot
of the shrinkage factors $\hat{\lambda}/\left(  \hat{\lambda}+A_{i}\right)  $
or $1-\hat{b}_{i}$ of all the shrinkage estimators for the case of all
players.}%
\label{fig.: Shrnkage factors for the baseball data}%
\end{figure}

Figure \ref{fig.: Shrnkage factors for the baseball data} shows the shrinkage
factors of all the shrinkage estimators with or without the covariates for the
all-players case of Table \ref{tab.: Baseball data prediction}. We see that
the shrinkage factors are all reduced after including the covariates. This
makes intuitive sense because the shrinkage location now contains the
covariate information, and each shrinkage estimator uses this information by
shrinking more toward it, resulting in smaller shrinkage
factors.

\section{Conclusion and Discussion}

\label{Sec.: Conclusion}

Inspired by the idea of unbiased risk estimate (URE) proposed in
\cite{Xie&Kou&Brown-2012}, we extend the URE framework to multivariate
heteroscedastic linear models, which are more realistic in practical
applications, especially for regression data that exhibits heteroscedasticity.
Several parallel URE shrinkage estimators in the regression case are proposed,
and these URE shrinkage estimators are all asymptotically optimal in risk
compared to other shrinkage estimators, including the classical empirical
Bayes ones. We also propose semiparametric estimators and conduct simulation
to assess their performance under both normal and non-normal data. For data
sets that exhibit a good linear relationship between the covariates and the
response, a semiparametric URE estimator is expected to provide good
estimation result, as we saw in the baseball data. It is also worth
emphasizing that the risk optimality for the parametric and semiparametric URE
estimators does not depend on the normality assumption of the sampling
distribution of $Y_{i}$.

We conclude this article by extending the main results to the case of weighted
mean squared error loss.

\textbf{Weighted mean squared error loss.} One might want to consider the more
general \emph{weighted mean squared error} as the loss function:%
\[
l_{p}\left(  \boldsymbol{\theta},\boldsymbol{\hat{\theta}};\boldsymbol{\psi
}\right)  =\dfrac{1}{p}\sum_{i=1}^{p}\psi_{i}\left(  \theta_{i}-\hat{\theta
}_{i}\right)  ^{2},
\]
where $\psi_{i}>0$ are known weights such that $\sum_{i=1}^{p}\psi_{i}=p$. The
framework proposed in this article is straightforward to generalize to this case.

For Model II, we only need to study the equivalent problem by the following
transformation%
\begin{equation}
Y_{i}\rightarrow\sqrt{\psi_{i}}Y_{i},\ \theta_{i}\rightarrow\sqrt{\psi_{i}%
}\theta_{i},\ \boldsymbol{X}_{i}\rightarrow\sqrt{\psi_{i}}\boldsymbol{X}%
_{i},\ A_{i}\rightarrow\psi_{i}A_{i}, \label{Weighted Loss transformation}%
\end{equation}
and restate the corresponding regularity conditions in Theorem
\ref{Thm.: Main Theorem} by the transformed data and parameters. We then
reduce the weighted mean square error problem back to the same setting we
study in this article under the classical loss function (\ref{Loss Function}).

Model I is more sophisticated than Model II to generalize. In addition to the
transformation in equation (\ref{Weighted Loss transformation}), we also need
$\lambda\rightarrow\psi_{i}\lambda$ in every term related to the individual
unit $i$. Thus,%
\[
\sqrt{\psi_{i}}\theta_{i}|\boldsymbol{X},\boldsymbol{\beta},\lambda
\overset{\text{indep.}}{\sim}N\left(  \sqrt{\psi_{i}}\boldsymbol{X}_{i}%
^{T}\boldsymbol{\beta},\lambda\psi_{i}\right)  ,
\]
so these transformed parameters $\sqrt{\psi_{i}}\theta_{i}$ are also
heteroscedastic in the sense that they have different weights, while the
setting we study before assumes all the weights on the $\theta_{i}$ are one.
However, if we carefully examine the proof of Theorem \ref{Thm.: Main Theorem}
for the case of Model I, we can see that actually we do not much require the
equal weights on the $\theta_{i}$'s. What is important in the proof is that
the shrinkage factor for unit $i$ is always of the form $A_{i}/\left(
A_{i}+\lambda\right)  $, which is \emph{invariant} under the transformation
$A_{i}\rightarrow\psi_{i}A_{i}$ and $\lambda\rightarrow\psi_{i}\lambda$. Thus,
after reformulating the regularity conditions in Theorem
\ref{Thm.: Main Theorem} by the transformed data and parameters, we can still
follow the same proof to conclude the risk optimality of URE estimators
(parametric or semiparametric) even under the consideration of weighted mean
squared error loss.

For completeness, here we state the most general result under the
semiparametric setting for Model I. Let%
\begin{align*}
\boldsymbol{\hat{\theta}}_{SP,\boldsymbol{\psi}}^{\mathrm{URE}}  &  =\left(
\boldsymbol{I}_{p}-\mathrm{diag}\left(  \boldsymbol{\hat{b}}_{\boldsymbol{\psi
}}^{\mathrm{URE}}\right)  \right)  \boldsymbol{Y}+\mathrm{diag}\left(
\boldsymbol{\hat{b}}_{\boldsymbol{\psi}}^{\mathrm{URE}}\right)
\boldsymbol{\hat{\mu}}_{\boldsymbol{\psi}}^{\mathrm{URE}},\\
\mathrm{URE}\left(  \boldsymbol{b},\boldsymbol{\mu};\boldsymbol{\psi}\right)
&  =\dfrac{1}{p}\sum_{i=1}^{p}\psi_{i}\left(  b_{i}^{2}\left(  Y_{i}-\mu
_{i}\right)  ^{2}+\left(  1-2b_{i}\right)  A_{i}\right)  ,\\
\left(  \boldsymbol{\hat{b}}_{\boldsymbol{\psi}}^{\mathrm{URE}}%
,\boldsymbol{\hat{\mu}}_{\boldsymbol{\psi}}^{\mathrm{URE}}\right)   &
=\operatorname*{argmin}\limits_{\boldsymbol{b}\in\mathrm{MON}\left(
\boldsymbol{A}\right)  ,\;\boldsymbol{\mu}\in\mathcal{L}_{\mathrm{row}}\left(
\boldsymbol{X}\right)  }\mathrm{URE}\left(  \boldsymbol{b},\boldsymbol{\mu
};\boldsymbol{\psi}\right)  .
\end{align*}

\begin{theorem}
Assume the following five conditions $\left(  \psi\text{-}\mathrm{A}\right)  $
$\sum_{i=1}^{p}\psi_{i}^{2}A_{i}^{2}=O\left(  p\right)  $, $\left(
\psi\text{-}\mathrm{B}\right)  $ $\sum_{i=1}^{p}\psi_{i}^{2}A_{i}\theta
_{i}^{2}=O\left(  p\right)  $, $\left(  \psi\text{-}\mathrm{C}\right)  $
$\sum_{i=1}^{p}\psi_{i}\theta_{i}^{2}=O\left(  p\right)  $, $\left(
\psi\text{-}\mathrm{D}\right)  $ $p^{-1}\sum_{i=1}^{p}\psi_{i}^{2}%
A_{i}\boldsymbol{X}_{i}\boldsymbol{X}_{i}^{T}$ converges, and $\left(
\psi\text{-}\mathrm{E}\right)  $ $p^{-1}\sum_{i=1}^{p}\psi_{i}\boldsymbol{X}%
_{i}\boldsymbol{X}_{i}^{T}\rightarrow\boldsymbol{\Omega}_{\boldsymbol{\psi}}>0$. Then we have%
\[
\sup\limits_{\boldsymbol{b}\in\mathrm{MON}\left(  \boldsymbol{A}\right)
,\;\boldsymbol{\mu}\in\mathcal{L}_{\boldsymbol{\psi}}}\left\vert
\mathrm{URE}\left(  \boldsymbol{b},\boldsymbol{\mu};\boldsymbol{\psi}\right)
-l_{p}\left(  \boldsymbol{\theta},\boldsymbol{\hat{\theta}}^{\boldsymbol{b}%
,\boldsymbol{\mu}};\boldsymbol{\psi}\right)  \right\vert
\underset{p\rightarrow\infty}{\rightarrow}0\text{ in }L^{1}\text{,}%
\]
where $\boldsymbol{\mu}\in\mathcal{L}_{\boldsymbol{\psi}}$ if and only if
$\boldsymbol{\mu}\in\mathcal{L}_{\mathrm{row}}\left(  \boldsymbol{X}\right)  $
and%
\[
\sum_{i=1}^{p}\psi_{i}\mu_{i}^{2}\leq Mp^{\kappa}\sum_{i=1}^{p}\psi_{i}%
Y_{i}^{2}%
\]
for a large and fixed constant $M$ and a fixed exponent $\kappa \in \left[ 0, 1/2 \right) $. As
a corollary, for any estimator $\boldsymbol{\hat{\theta}}^{\boldsymbol{\hat
{b}}_{p},\boldsymbol{\hat{\mu}}_{p}}= (\boldsymbol{I}_{p}-\mathrm{diag}%
( \boldsymbol{\hat{b}}_{p} ) ) \boldsymbol{Y}+\mathrm{diag} ( \boldsymbol{\hat{b}%
}_{p} ) \boldsymbol{\hat{\mu}}_{p}$ with $\boldsymbol{\hat{b}}_{p}%
\in\mathrm{MON}\left(  \boldsymbol{A}\right)  $ and $\boldsymbol{\hat{\mu}%
}_{p}\in\mathcal{L}_{\boldsymbol{\psi}}$, we have%
\begin{align*}
\lim\limits_{p\rightarrow\infty}\mathbb{P}\left(  l_{p}\left(
\boldsymbol{\theta},\boldsymbol{\hat{\theta}}_{SP,\boldsymbol{\psi}%
}^{\mathrm{URE}}\right)  \geq l_{p}\left(  \boldsymbol{\theta}%
,\boldsymbol{\hat{\theta}}^{\boldsymbol{\hat{b}}_{p},\boldsymbol{\hat{\mu}%
}_{p}}\right)  +\epsilon\right)   &  =0\ \ \ \ \forall\epsilon>0,\\
\limsup\limits_{p\rightarrow\infty}\left(  R_{p}\left(  \boldsymbol{\theta
},\boldsymbol{\hat{\theta}}_{SP,\boldsymbol{\psi}}^{\mathrm{URE}}\right)
-R_{p}\left(  \boldsymbol{\theta},\boldsymbol{\hat{\theta}}^{\boldsymbol{\hat
{b}}_{p},\boldsymbol{\hat{\mu}}_{p}}\right)  \right)   &  \leq0.
\end{align*}
\end{theorem}

\section*{Appendix: Proofs and Derivations}

\addcontentsline{toc}{section}{Appendix}

\textbf{Proof of Lemma \ref{Lma.: Generic posterior dist. and marginal}.} We
can write $\boldsymbol{\theta}=\boldsymbol{\mu}+\boldsymbol{Z}_{1}$ and
$\boldsymbol{Y}=\boldsymbol{\theta}+\boldsymbol{Z}_{2}$, where $\boldsymbol{Z}%
_{1}\sim\mathcal{N}_{p}(\boldsymbol{0},\boldsymbol{B})$ and $\boldsymbol{Z}%
_{2}\sim\mathcal{N}_{p}(\boldsymbol{0},\boldsymbol{A})$ are independent.
Jointly $\dbinom{\boldsymbol{Y}}{\boldsymbol{\theta}}$ is still multivariate
normal with mean vector $\dbinom{\boldsymbol{\mu}}{\boldsymbol{\mu}}$ and
covariance matrix $\left(
\begin{array}
[c]{cc}%
\boldsymbol{A}+\boldsymbol{B} & \boldsymbol{B}\\
\boldsymbol{B} & \boldsymbol{B}%
\end{array}
\right)  $. The result follows immediately from the conditional distribution
of a multivariate normal distribution.

\textbf{Proof of Theorem \ref{Thm.: Main Theorem}.} We start from decomposing
the difference between the URE and the actual loss as%
\begin{align}
&  \mathrm{URE}\left(  \boldsymbol{B},\boldsymbol{\mu}\right)  -l_{p}\left(
\boldsymbol{\theta},\boldsymbol{\hat{\theta}}^{\boldsymbol{B},\boldsymbol{\mu
}}\right) \nonumber\\
&  =\mathrm{URE}\left(  \boldsymbol{B},\boldsymbol{0}_{p}\right)
-l_{p}\left(  \boldsymbol{\theta},\boldsymbol{\hat{\theta}}^{\boldsymbol{B}%
,\boldsymbol{0}_{p}}\right)  -\dfrac{2}{p}\mathrm{tr}\left(  \boldsymbol{A}%
\left(  \boldsymbol{A}+\boldsymbol{B}\right)  ^{-1}\boldsymbol{\mu}\left(
\boldsymbol{Y}-\boldsymbol{\theta}\right)  ^{T}\right) \label{firsteq}\\
&  =\dfrac{1}{p}\mathrm{tr}\left(  \boldsymbol{YY}^{T}-\boldsymbol{A}%
-\boldsymbol{\theta\theta}^{T}\right)  -\dfrac{2}{p}\mathrm{tr}\left(
\boldsymbol{B}\left(  \boldsymbol{A}+\boldsymbol{B}\right)  ^{-1}\left(
\boldsymbol{YY}^{T}-\boldsymbol{Y\theta}^{T}-\boldsymbol{A}\right)  \right)
\nonumber\\
&  -\dfrac{2}{p}\mathrm{tr}\left(  \boldsymbol{A}\left(  \boldsymbol{A}%
+\boldsymbol{B}\right)  ^{-1}\boldsymbol{\mu}\left(  \boldsymbol{Y}%
-\boldsymbol{\theta}\right)  ^{T}\right) \label{2ndeq}\\
&  =\left(  \mathrm{I}\right)  +\left(  \mathrm{II}\right)  +\left(
\mathrm{III}\right)  .\nonumber
\end{align}
To verify the first equality (\ref{firsteq}), note that%
\begin{align*}
&  \mathrm{URE}\left(  \boldsymbol{B},\boldsymbol{\mu}\right)  -\mathrm{URE}%
\left(  \boldsymbol{B},\boldsymbol{0}_{p}\right) \\
&  =\dfrac{1}{p}\left\Vert \boldsymbol{A}\left(  \boldsymbol{A}+\boldsymbol{B}%
\right)  ^{-1}\left(  \boldsymbol{Y}-\boldsymbol{\mu}\right)  \right\Vert
^{2}-\dfrac{1}{p}\left\Vert \boldsymbol{A}\left(  \boldsymbol{A}%
+\boldsymbol{B}\right)  ^{-1}\boldsymbol{Y}\right\Vert ^{2}\\
&  =-\dfrac{1}{p}\mathrm{tr}\left(  \boldsymbol{\mu}^{T}\left(  \boldsymbol{A}%
\left(  \boldsymbol{A}+\boldsymbol{B}\right)  ^{-1}\right)  ^{T}%
\boldsymbol{A}\left(  \boldsymbol{A}+\boldsymbol{B}\right)  ^{-1}\left(
2\boldsymbol{Y}-\boldsymbol{\mu}\right)  \right)  ,\\
&  l_{p}\left(  \boldsymbol{\theta},\boldsymbol{\hat{\theta}}^{\boldsymbol{B}%
,\boldsymbol{\mu}}\right)  -l_{p}\left(  \boldsymbol{\theta},\boldsymbol{\hat
{\theta}}^{\boldsymbol{B},\boldsymbol{0}_{p}}\right) \\
&  =\dfrac{1}{p}\left\Vert \left(  \boldsymbol{I}_{p}-\boldsymbol{A}\left(
\boldsymbol{A}+\boldsymbol{B}\right)  ^{-1}\right)  \boldsymbol{Y}%
+\boldsymbol{A}\left(  \boldsymbol{A}+\boldsymbol{B}\right)  ^{-1}%
\boldsymbol{\mu}-\boldsymbol{\theta}\right\Vert ^{2}-\dfrac{1}{p}\left\Vert
\left(  \boldsymbol{I}_{p}-\boldsymbol{A}\left(  \boldsymbol{A}+\boldsymbol{B}%
\right)  ^{-1}\right)  \boldsymbol{Y}-\boldsymbol{\theta}\right\Vert ^{2}\\
&  =\dfrac{1}{p}\mathrm{tr}\left(  \boldsymbol{\mu}^{T}\left(  \boldsymbol{A}%
\left(  \boldsymbol{A}+\boldsymbol{B}\right)  ^{-1}\right)  ^{T}\left(
2\left(  \left(  \boldsymbol{I}_{p}-\boldsymbol{A}\left(  \boldsymbol{A}%
+\boldsymbol{B}\right)  ^{-1}\right)  \boldsymbol{Y-\theta}\right)
+\boldsymbol{A}\left(  \boldsymbol{A}+\boldsymbol{B}\right)  ^{-1}%
\boldsymbol{\mu}\right)  \right)  .
\end{align*}
(\ref{firsteq}) then follows by rearranging the terms. To verify the second
equality (\ref{2ndeq}), note%
\begin{align*}
&  \mathrm{URE}\left(  \boldsymbol{B},\boldsymbol{0}_{p}\right)  -l_{p}\left(
\boldsymbol{\theta},\boldsymbol{\hat{\theta}}^{\boldsymbol{B},\boldsymbol{0}%
_{p} }\right) \\
&  =\dfrac{1}{p}\left\Vert \boldsymbol{A}\left(  \boldsymbol{A}+\boldsymbol{B}
\right)  ^{-1}\boldsymbol{Y}\right\Vert ^{2}-\dfrac{1}{p}\left\Vert \left(
\boldsymbol{I}_{p}-\boldsymbol{A}\left(  \boldsymbol{A}+\boldsymbol{B}\right)
^{-1}\right)  \boldsymbol{Y}-\boldsymbol{\theta}\right\Vert ^{2}\\
&  +\dfrac{1}{p}\mathrm{tr}\left(  \boldsymbol{A}-2\boldsymbol{A}\left(
\boldsymbol{A}+\boldsymbol{B}\right)  ^{-1}\boldsymbol{A}\right) \\
&  =\dfrac{1}{p}\mathrm{tr}\left(  \left(  \boldsymbol{Y}-2\left(
\boldsymbol{I}_{p}-\boldsymbol{A}\left(  \boldsymbol{A}+\boldsymbol{B}\right)
^{-1}\right)  \boldsymbol{Y}+\boldsymbol{\theta}\right)  ^{T}\left(
\boldsymbol{Y}-\boldsymbol{\theta}\right)  \right)  +\dfrac{1}{p}
\mathrm{tr}\left(  \boldsymbol{A}-2\boldsymbol{A}\left(  \boldsymbol{A}
+\boldsymbol{B}\right)  ^{-1}\boldsymbol{A}\right) \\
&  =\dfrac{1}{p}\mathrm{tr}\left(  \boldsymbol{YY}^{T}-\boldsymbol{A}
-\boldsymbol{\theta\theta}^{T}\right)  -\dfrac{2}{p}\mathrm{tr}\left(
\boldsymbol{B}\left(  \boldsymbol{A}+\boldsymbol{B}\right)  ^{-1}\left(
\boldsymbol{Y}\left(  \boldsymbol{Y}-\boldsymbol{\theta}\right)
^{T}-\boldsymbol{A}\right)  \right)  .
\end{align*}

With the decomposition, we want to prove separately the uniform $L^{1}$
convergence of the three terms $\left(  \mathrm{I}\right)  $, $\left(
\mathrm{II}\right)  $, and $\left(  \mathrm{III}\right)  $.

\textit{Proof for the case of Model I.}

The uniform $L^{2}$ convergence of $\left(  \mathrm{I}\right)  $ and $\left(
\mathrm{II}\right)  $ has been shown in Theorem 3.1 of
\cite{Xie&Kou&Brown-2012} under our assumptions $\left(  \mathrm{A}\right)  $
and $\left(  \mathrm{B}\right)  $, so we focus on $\left(  \mathrm{III}%
\right)  $, i.e., we want to show that $\sup\limits_{0\leq\lambda\leq
\infty,\;\boldsymbol{\mu}\in\mathcal{L}}\left\vert \left(  \mathrm{III}%
\right)  \right\vert \rightarrow0$ in $L^{1}$ as $p\rightarrow\infty$.

Without loss of generality, let us assume $A_{1}\leq A_{2}\leq\cdots\leq
A_{p}$. We have
\begin{align*}
\sup\limits_{0\leq\lambda\leq\infty,\;\boldsymbol{\mu}\in\mathcal{L}%
}\left\vert \left(  \mathrm{III}\right)  \right\vert  &  =\dfrac{2}{p}%
\sup\limits_{0\leq\lambda\leq\infty,\;\boldsymbol{\mu}\in\mathcal{L}%
}\left\vert \sum_{i=1}^{p}\dfrac{A_{i}}{A_{i}+\lambda}\mu_{i}\left(
Y_{i}-\theta_{i}\right)  \right\vert \\
&  \leq\dfrac{2}{p}\sup\limits_{\boldsymbol{\mu}\in\mathcal{L}}\sup
\limits_{0\leq c_{1}\leq\cdots\leq c_{p}\leq1}\left\vert \sum_{i=1}^{p}%
c_{i}\mu_{i}\left(  Y_{i}-\theta_{i}\right)  \right\vert =\dfrac{2}{p}%
\sup\limits_{\boldsymbol{\mu}\in\mathcal{L}}\max\limits_{1\leq j\leq
p}\left\vert \sum_{i=j}^{p}\mu_{i}\left(  Y_{i}-\theta_{i}\right)  \right\vert
,
\end{align*}
where the last equality follows from Lemma 2.1 of \cite{Li-1986}. For a
generic $p$-dimensional vector $\boldsymbol{v}$, we denote $[\boldsymbol{v}%
]_{j:p}=(0,\ldots0,v_{j},v_{j+1},\ldots,v_{p})$. Let $\boldsymbol{P}%
_{\boldsymbol{X}}=\boldsymbol{X}^{T}\left(  \boldsymbol{XX}^{T}\right)
^{-1}\boldsymbol{X}$ be the projection matrix onto $\mathcal{L}_{\mathrm{row}%
}\left(  \boldsymbol{X}\right)  $. Then since $\mathcal{L}\subset
\mathcal{L}_{\mathrm{row}}\left(  \boldsymbol{X}\right)  $, we have%
\begin{align*}
&  \dfrac{2}{p}\sup\limits_{\boldsymbol{\mu}\in\mathcal{L}}\max\limits_{1\leq
j\leq p}\left\vert \sum_{i=j}^{p}\mu_{i}\left(  Y_{i}-\theta_{i}\right)
\right\vert =\dfrac{2}{p}\max\limits_{1\leq j\leq p}\sup
\limits_{\boldsymbol{\mu}\in\mathcal{L}}\left\vert \boldsymbol{\mu}%
^{T}[\boldsymbol{Y}-\boldsymbol{\theta}]_{j:p}\right\vert \\
=  &  \dfrac{2}{p}\max\limits_{1\leq j\leq p}\sup\limits_{\boldsymbol{\mu}%
\in\mathcal{L}}\left\vert \boldsymbol{\mu}^{T}\boldsymbol{P}_{\boldsymbol{X}%
}[\boldsymbol{Y}-\boldsymbol{\theta}]_{j:p}\right\vert \leq\dfrac{2}{p}%
\max\limits_{1\leq j\leq p}\sup\limits_{\boldsymbol{\mu}\in\mathcal{L}%
}\left\Vert \boldsymbol{\mu}\right\Vert \times\left\Vert \boldsymbol{P}%
_{\boldsymbol{X}}[\boldsymbol{Y}-\boldsymbol{\theta}]_{j:p}\right\Vert \\
= &  \dfrac{2}{p}\max\limits_{1\leq j\leq p}Mp^{\kappa}\left\Vert
\boldsymbol{Y}\right\Vert \times\left\Vert \boldsymbol{P}_{\boldsymbol{X}%
}[\boldsymbol{Y}-\boldsymbol{\theta}]_{j:p}\right\Vert .
\end{align*}
Cauchy-Schwarz inequality thus gives%
\begin{equation}
\mathbb{E}\left(  \sup\limits_{0\leq\lambda\leq\infty,\boldsymbol{\mu}%
\in\mathcal{L}}\left\vert \left(  \mathrm{III}\right)  \right\vert \right)
\leq2Mp^{\kappa-1}\sqrt{\mathbb{E}\left(  \left\Vert \boldsymbol{Y}\right\Vert
^{2}\right)  }\times\sqrt{\mathbb{E}\left(  \max\limits_{1\leq j\leq
p}\left\Vert \boldsymbol{P}_{\boldsymbol{X}}[\boldsymbol{Y}-\boldsymbol{\theta
}]_{j:p}\right\Vert ^{2}\right)  }. \label{Esup}%
\end{equation}
It is straightforward to see that, by conditions (A) and (C),%
\[
\sqrt{\mathbb{E}\left(  \left\Vert \boldsymbol{Y}\right\Vert ^{2}\right)
}=\sqrt{\mathbb{E(}\sum\nolimits_{i=1}^{p}Y_{i}^{2})}=\sqrt{\sum
\nolimits_{i=1}^{p}\left(  \theta_{i}^{2}+A_{i}\right)  }=O\left(
p^{1/2}\right)  .
\]
For the second term on the right hand side of (\ref{Esup}), let
$\boldsymbol{P}_{\boldsymbol{X}}=\boldsymbol{\Gamma D\Gamma}^{T}$ denote the
spectral decomposition. Clearly,%
\[
\boldsymbol{D}=\mathrm{diag}\left(  \underset{k\text{ copies}%
}{\underbrace{1,...,1}},\underset{p-k\text{ copies}}{\underbrace{0,...,0}%
}\right)  .
\]
It follows that%
\begin{align*}
&  \mathbb{E}\left(  \max\limits_{1\leq j\leq p}\left\Vert \boldsymbol{P}%
_{\boldsymbol{X}}[\boldsymbol{Y}-\boldsymbol{\theta}]_{j:p}\right\Vert
^{2}\right)  =\mathbb{E}\left(  \max\limits_{1\leq j\leq p}[\boldsymbol{Y}%
-\boldsymbol{\theta}]_{j:p}^{T}\boldsymbol{P}_{\boldsymbol{X}}[\boldsymbol{Y}%
-\boldsymbol{\theta}]_{j:p}\right) \\
&  =\mathbb{E}\left(  \max\limits_{1\leq j\leq p}\mathrm{tr}\left(
\boldsymbol{D\Gamma}^{T}[\boldsymbol{Y}-\boldsymbol{\theta}]_{j:p}\left(
\boldsymbol{\Gamma}^{T}[\boldsymbol{Y}-\boldsymbol{\theta}]_{j:p}\right)
^{T}\right)  \right)  =\mathbb{E}\left(  \max\limits_{1\leq j\leq p}\sum
_{l=1}^{k}\left[  \boldsymbol{\Gamma}^{T}[\boldsymbol{Y}-\boldsymbol{\theta
}]_{j:p}\right]  _{l}^{2}\right) \\
&  =\mathbb{E}\left(  \max\limits_{1\leq j\leq p}\sum_{l=1}^{k}\left(
\sum_{m=j}^{p}\left[  \boldsymbol{\Gamma}^{T}\right]  _{lm}\left(
Y_{m}-\theta_{m}\right)  \right)  ^{2}\right) \\
&  \leq\mathbb{E}\left(  \sum_{l=1}^{k}\max\limits_{1\leq j\leq p}\left(
\sum_{m=j}^{p}\left[  \boldsymbol{\Gamma}^{T}\right]  _{lm}\left(
Y_{m}-\theta_{m}\right)  \right)  ^{2}\right)  =\sum_{l=1}^{k}\mathbb{E}%
\left(  \max\limits_{1\leq j\leq p}\left(  \sum_{m=j}^{p}\left[
\boldsymbol{\Gamma}^{T}\right]  _{lm}\left(  Y_{m}-\theta_{m}\right)  \right)
^{2}\right)  .
\end{align*}
For each $l$, $M_{j}^{\left(  l\right)  }=\sum_{m=p-j+1}^{p}\left[
\boldsymbol{\Gamma}^{T}\right]  _{lm}\left(  Y_{m}-\theta_{m}\right)  $ forms
a martingale, so by Doob's $L^{p}$ maximum inequality,%
\begin{align*}
\mathbb{\mathbb{E}}\left(  \max\limits_{1\leq j\leq p}\left(  M_{j}^{\left(
l\right)  }\right)  ^{2}\right)   &  \leq4\mathbb{E}\left(  M_{p}^{\left(
l\right)  }\right)  ^{2}=4\mathbb{E}\left(  \sum_{m=1}^{p}\left[
\boldsymbol{\Gamma}^{T}\right]  _{lm}\left(  Y_{m}-\theta_{m}\right)  \right)
^{2}\\
&  =4\sum_{m=1}^{p}\left[  \boldsymbol{\Gamma}^{T}\right]  _{lm}^{2}%
A_{m}=4\left[  \boldsymbol{\Gamma}^{T}\boldsymbol{A\Gamma}\right]  _{ll}.
\end{align*}
Therefore,%
\begin{align*}
&  \mathbb{E}\left(  \max\limits_{1\leq j\leq p}\left\Vert \boldsymbol{P}%
_{\boldsymbol{X}}[\boldsymbol{Y}-\boldsymbol{\theta}]_{j:p}\right\Vert
^{2}\right)  \leq\sum_{l=1}^{k}4\left[  \boldsymbol{\Gamma}^{T}%
\boldsymbol{A\Gamma}\right]  _{ll}\\
&  =4\sum_{l=1}^{p}\left[  \boldsymbol{D}\right]  _{ll}\left[
\boldsymbol{\Gamma}^{T}\boldsymbol{A\Gamma}\right]  _{ll}=4\ \mathrm{tr}%
\left(  \boldsymbol{D\Gamma}^{T}\boldsymbol{A\Gamma}\right)  =4\ \mathrm{tr}%
\left(  \boldsymbol{P}_{\boldsymbol{X}}\boldsymbol{A}\right) \\
&  =4\ \mathrm{tr}\left(  \boldsymbol{X}^{T}\left(  \boldsymbol{XX}%
^{T}\right)  ^{-1}\boldsymbol{XA}\right)  =4\ \mathrm{tr}\left(  \left(
\boldsymbol{XX}^{T}\right)  ^{-1}\boldsymbol{XAX}^{T}\right)  =O\left(
1\right)  ,
\end{align*}
where the last equality uses conditions $\left(  \mathrm{D}\right)  $ and
$\left(  \mathrm{E}\right)  $. We finally obtain%
\[
\mathbb{E}\left(  \sup\limits_{0\leq\lambda\leq\infty,\;\boldsymbol{\mu}%
\in\mathcal{L}}\left\vert \left(  \mathrm{III}\right)  \right\vert \right)
\leq o\left(  p^{-1/2}\right)  \times O\left(  p^{1/2}\right)  \times O\left(
1\right)  =o\left(  1\right)  .
\]

\textit{Proof for the case of Model II.}

Under Model II, we know that%
\[
\sum_{i=1}^{p}A_{i}\theta_{i}^{2}=\boldsymbol{\theta}^{T}\boldsymbol{A\theta
}=\boldsymbol{\beta}^{T}(\boldsymbol{XAX}^{T})\boldsymbol{\beta}=O\left(
p\right)
\]
by condition $\left(  \mathrm{D}\right)  $. In other words, condition $\left(
\mathrm{D}\right)  $ implies condition $\left(  \mathrm{B}\right)  $.
Therefore, we know that the term $\left(  \mathrm{I}\right)  \rightarrow0$ in
$L^{2}$ as shown in Theorem 3.1 of \cite{Xie&Kou&Brown-2012}, and we only need
to show the uniform $L^{1}$ convergence of the other two terms, $\left(
\mathrm{II}\right)  $ and $\left(  \mathrm{III}\right)  $.

Recall that $\boldsymbol{B}\in\mathcal{B}=\left\{  \lambda\boldsymbol{X}%
^{T}\boldsymbol{WX}:\lambda>0\right\}  $ has only rank $k$ under Model II. We
can reexpress $\left(  \mathrm{II}\right)  $ and $\left(  \mathrm{III}\right)
$ in terms of low rank matrices. Let $\boldsymbol{V}=\left(  \boldsymbol{XA}%
^{-1}\boldsymbol{X}^{T}\right)  ^{-1}$. Woodbury formula gives%
\begin{align*}
\left(  \boldsymbol{A}+\boldsymbol{B}\right)  ^{-1}  &  =\left(
\boldsymbol{A}+\lambda\boldsymbol{X}^{T}\boldsymbol{WX}\right)  ^{-1}%
=\boldsymbol{A}^{-1}-\boldsymbol{A}^{-1}\lambda\boldsymbol{X}^{T}\left(
\boldsymbol{W}^{-1}+\lambda\boldsymbol{V}^{-1}\right)  ^{-1}\boldsymbol{XA}%
^{-1}\\
&  =\boldsymbol{A}^{-1}-\boldsymbol{A}^{-1}\lambda\boldsymbol{X}%
^{T}\boldsymbol{W}\left(  \lambda\boldsymbol{W}+\boldsymbol{V}\right)
^{-1}\boldsymbol{VXA}^{-1},
\end{align*}
which tells us
\[
\boldsymbol{B}\left(  \boldsymbol{A}+\boldsymbol{B}\right)  ^{-1}%
=\boldsymbol{I}_{p}-\boldsymbol{A}\left(  \boldsymbol{A}+\boldsymbol{B}%
\right)  ^{-1}=\lambda\boldsymbol{X}^{T}\boldsymbol{W}\left(  \lambda
\boldsymbol{W}+\boldsymbol{V}\right)  ^{-1}\boldsymbol{VXA}^{-1}.
\]
Let $\boldsymbol{U}\boldsymbol{\Lambda U}^{T}$ be the spectral decomposition
of $\boldsymbol{W}^{-1/2}\boldsymbol{VW}^{-1/2}$, i.e., $\boldsymbol{W}%
^{-1/2}\boldsymbol{VW}^{-1/2}=\boldsymbol{U\Lambda}\boldsymbol{U}^{T}$, where
$\boldsymbol{\Lambda}=\mathrm{diag}\left(  d_{1},...,d_{k}\right)  $ with
$d_{1}\leq\cdots\leq d_{k}$. Then $\left(  \lambda\boldsymbol{W}%
+\boldsymbol{V}\right)  ^{-1}=\boldsymbol{W}^{-1/2}\left(  \lambda
\boldsymbol{I}_{k}+\boldsymbol{W}^{-1/2}\boldsymbol{VW}^{-1/2}\right)
^{-1}\boldsymbol{W}^{-1/2}=\boldsymbol{W}^{-1/2}\boldsymbol{U}\left(
\lambda\boldsymbol{I}_{k}+\boldsymbol{\Lambda}\right)  ^{-1}\boldsymbol{U}%
^{T}\boldsymbol{W}^{-1/2}$, from which we obtain%
\[
\boldsymbol{B}\left(  \boldsymbol{A}+\boldsymbol{B}\right)  ^{-1}%
=\lambda\boldsymbol{X}^{T}\boldsymbol{W}\left(  \lambda\boldsymbol{W}%
+\boldsymbol{V}\right)  ^{-1}\boldsymbol{VXA}^{-1}=\lambda\boldsymbol{X}%
^{T}\boldsymbol{W}^{1/2}\boldsymbol{U}\left(  \lambda\boldsymbol{I}%
_{k}+\boldsymbol{\Lambda}\right)  ^{-1}\boldsymbol{\Lambda U}^{T}%
\boldsymbol{W}^{1/2}\boldsymbol{XA}^{-1}.
\]
If we denote $\boldsymbol{Z}=\boldsymbol{U}^{T}\boldsymbol{W}^{1/2}%
\boldsymbol{X}$, i.e., $\boldsymbol{Z}$ is the transformed covariate matrix,
then $\boldsymbol{B}\left(  \boldsymbol{A}+\boldsymbol{B}\right)
^{-1}=\lambda\boldsymbol{Z}^{T}\left(  \lambda\boldsymbol{I}_{k}%
+\boldsymbol{\Lambda}\right)  ^{-1}\boldsymbol{\Lambda}\boldsymbol{Z}%
\boldsymbol{A}^{-1}$. It follows that
\begin{align*}
\left(  \mathrm{II}\right)   &  =-\dfrac{2}{p}\mathrm{tr}\left(
\boldsymbol{B}\left(  \boldsymbol{A}+\boldsymbol{B}\right)  ^{-1}\left(
\boldsymbol{YY}^{T}-\boldsymbol{Y\theta}^{T}-\boldsymbol{A}\right)  \right) \\
&  =-\dfrac{2}{p}\mathrm{tr}\left(  \lambda\boldsymbol{Z}^{T}\left(
\lambda\boldsymbol{I}_{k}+\boldsymbol{\Lambda}\right)  ^{-1}%
\boldsymbol{\Lambda}\boldsymbol{Z}\boldsymbol{A}^{-1}\left(  \boldsymbol{YY}%
^{T}-\boldsymbol{Y\theta}^{T}-\boldsymbol{A}\right)  \right) \\
&  =-\dfrac{2}{p}\mathrm{tr}\left(  \lambda\left(  \lambda\boldsymbol{I}%
_{k}+\boldsymbol{\Lambda}\right)  ^{-1}\boldsymbol{\Lambda}\boldsymbol{Z}%
\boldsymbol{A}^{-1}\left(  \boldsymbol{YY}^{T}-\boldsymbol{Y\theta}%
^{T}-\boldsymbol{A}\right)  \boldsymbol{Z}^{T}\right)  ,\\
\left(  \mathrm{III}\right)   &  =-\dfrac{2}{p}\mathrm{tr}\left(
\boldsymbol{A}\left(  \boldsymbol{A}+\boldsymbol{B}\right)  ^{-1}%
\boldsymbol{\mu}\left(  \boldsymbol{Y}-\boldsymbol{\theta}\right)  ^{T}\right)
\\
&  =-\dfrac{2}{p}\mathrm{tr}\left(  \left(  \boldsymbol{I}_{p}-\lambda
\boldsymbol{Z}^{T}\left(  \lambda\boldsymbol{I}_{k}+\boldsymbol{\Lambda
}\right)  ^{-1}\boldsymbol{\Lambda}\boldsymbol{Z}\boldsymbol{A}^{-1}\right)
\boldsymbol{\mu}\left(  \boldsymbol{Y}-\boldsymbol{\theta}\right)  ^{T}\right)
\\
&  =-\dfrac{2}{p}\mathrm{tr}\left(  \boldsymbol{\mu}\left(  \boldsymbol{Y}%
-\boldsymbol{\theta}\right)  ^{T}\right)  +\dfrac{2}{p}\mathrm{tr}\left(
\lambda\left(  \lambda\boldsymbol{I}_{k}+\boldsymbol{\Lambda}\right)
^{-1}\boldsymbol{\Lambda}\boldsymbol{Z}\boldsymbol{A}^{-1}\boldsymbol{\mu
}\left(  \boldsymbol{Y}-\boldsymbol{\theta}\right)  ^{T}\boldsymbol{Z}%
^{T}\right) \\
&  =\left(  \mathrm{III}\right)  _{1}+\left(  \mathrm{III}\right)  _{2}.
\end{align*}
We will next show that $\left(  \mathrm{II}\right)  $, $\left(  \mathrm{III}%
\right)  _{1}$, and $\left(  \mathrm{III}\right)  _{2}$ all uniformly converge
to zero in $L^{1}$, which will then complete our proof.

Let $\boldsymbol{\Xi}=\boldsymbol{ZA}^{-1}\left(  \boldsymbol{YY}%
^{T}-\boldsymbol{Y\theta}^{T}-\boldsymbol{A}\right)  \boldsymbol{Z}^{T}$. Then%
\begin{align*}
\sup\limits_{0\leq\lambda\leq\infty}\left\vert \left(  \mathrm{II}\right)
\right\vert  &  =\dfrac{2}{p}\sup\limits_{0\leq\lambda\leq\infty}\left\vert
\sum_{i=1}^{k}\dfrac{\lambda d_{i}}{\lambda+d_{i}}\left[  \boldsymbol{\Xi
}\right]  _{ii}\right\vert \\
&  \leq\dfrac{2}{p}\sup\limits_{0\leq c_{1}\leq\cdots\leq c_{k}\leq d_{k}%
}\left\vert \sum_{i=1}^{k}c_{i}\left[  \boldsymbol{\Xi}\right]  _{ii}%
\right\vert =\dfrac{2}{p}\max\limits_{1\leq j\leq k}\left\vert \sum_{i=j}%
^{k}d_{k}\left[  \boldsymbol{\Xi}\right]  _{ii}\right\vert ,
\end{align*}
where the last equality follows as in Lemma 2.1 of \cite{Li-1986}. As there
are finite number of terms in the summation and the maximization, it suffices
to show that%
\[
d_{k}\left[  \boldsymbol{\Xi}\right]  _{ii}/p\rightarrow0\text{ in }%
L^{2}\ \ \ \ \text{for all }1\leq i\leq k.
\]
To establish this, we note that $\left[  \boldsymbol{\Xi}\right]  _{ii}%
=\sum_{n=1}^{p}\sum_{m=1}^{p}\left(  A_{n}^{-1}Y_{n}\left(  Y_{m}-\theta
_{m}\right)  -\delta_{nm}\right)  \left[  \boldsymbol{Z}\right]  _{in}\left[
\boldsymbol{Z}\right]  _{im}$,%
\begin{align*}
\mathbb{E}\left(  \left[  \boldsymbol{\Xi}\right]  _{ii}^{2}\right)   &
=\sum_{n,m,n^{\prime},m^{\prime}}\mathbb{E}\left(  \left(  A_{n}^{-1}%
Y_{n}\left(  Y_{m}-\theta_{m}\right)  -\delta_{nm}\right)  \left(
A_{n^{\prime}}^{-1}Y_{n^{\prime}}\left(  Y_{m^{\prime}}-\theta_{m^{\prime}%
}\right)  -\delta_{n^{\prime}m^{\prime}}\right)  \right) \\
&  \times\left[  \boldsymbol{Z}\right]  _{in}\left[  \boldsymbol{Z}\right]
_{im}\left[  \boldsymbol{Z}\right]  _{in^{\prime}}\left[  \boldsymbol{Z}%
\right]  _{im^{\prime}}.
\end{align*}
Depending on $n,m,n^{\prime},m^{\prime}$ taking the same or distinct values,
we can break the summation into 15 disjoint cases:%
\begin{align*}
&  \sum_{\text{all distinct}}+\sum_{\text{three distinct, }n=m}+\sum
_{\text{three distinct, }n=n^{\prime}}+\sum_{\text{three distinct,
}n=m^{\prime}}\\
&  +\sum_{\text{three distinct, }m=n^{\prime}}+\sum_{\text{three distinct,
}m=m^{\prime}}+\sum_{\text{three distinct, }n^{\prime}=m^{\prime}}%
+\sum_{\text{two distinct, }n=m\text{, }n^{\prime}=m^{\prime}}\\
&  +\sum_{\text{two distinct, }n=n^{\prime}\text{, }m=m^{\prime}}%
+\sum_{\text{two distinct, }n=m^{\prime}\text{, }n^{\prime}=m}+\sum_{\text{two
distinct, }n=m=n^{\prime}}+\sum_{\text{two distinct, }n=m=m^{\prime}}\\
&  +\sum_{\text{two distinct, }n=n^{\prime}=m^{\prime}}+\sum_{\text{two
distinct, }m=n^{\prime}=m^{\prime}}+\sum_{n=m=n^{\prime}=m^{\prime}}.
\end{align*}
Many terms are zero. Straightforward evaluation of each summation gives%
\begin{align*}
\mathbb{E}\left(  \left[  \boldsymbol{\Xi}\right]  _{ii}^{2}\right)   &
=\sum_{n=1}^{p}\mathbb{E}\left(  \left(  A_{n}^{-1}Y_{n}\left(  Y_{n}%
-\theta_{n}\right)  -1\right)  ^{2}\right)  \left[  \boldsymbol{Z}\right]
_{in}^{4}\\
&  +\sum_{n=1}^{p}\sum_{m\neq n}\mathbb{E}\left(  \left(  A_{n}^{-1}%
Y_{n}\left(  Y_{m}-\theta_{m}\right)  \right)  ^{2}\right)  \left[
\boldsymbol{Z}\right]  _{in}^{2}\left[  \boldsymbol{Z}\right]  _{im}^{2}\\
&  +\sum_{n=1}^{p}\sum_{m\neq n}\mathbb{E}\left(  \left(  A_{n}^{-1}%
Y_{n}\left(  Y_{m}-\theta_{m}\right)  \right)  \left(  A_{m}^{-1}Y_{m}\left(
Y_{n}-\theta_{n}\right)  \right)  \right)  \left[  \boldsymbol{Z}\right]
_{in}^{2}\left[  \boldsymbol{Z}\right]  _{im}^{2}\\
&  +2\sum_{n=1}^{p}\sum_{m\neq n}\mathbb{E}\left(  \left(  A_{n}^{-1}%
Y_{n}\left(  Y_{n}-\theta_{n}\right)  -1\right)  \left(  A_{m}^{-1}%
Y_{m}\left(  Y_{n}-\theta_{n}\right)  \right)  \right)  \left[  \boldsymbol{Z}%
\right]  _{in}^{3}\left[  \boldsymbol{Z}\right]  _{im}\\
&  +\sum_{n=1}^{p}\sum_{m\neq n^{\prime},n^{\prime}\neq n,m\neq n}%
\mathbb{E}\left(  \left(  A_{m}^{-1}Y_{m}\left(  Y_{n}-\theta_{n}\right)
\right)  \left(  A_{n^{\prime}}^{-1}Y_{n^{\prime}}\left(  Y_{n}-\theta
_{n}\right)  \right)  \right)  \left[  \boldsymbol{Z}\right]  _{in}^{2}\left[
\boldsymbol{Z}\right]  _{im}\left[  \boldsymbol{Z}\right]  _{in^{\prime}}\\
&  =\sum_{n=1}^{p}\dfrac{2A_{n}+\theta_{n}^{2}}{A_{n}}\left[  \boldsymbol{Z}%
\right]  _{in}^{4}+\sum_{n=1}^{p}\sum_{m\neq n}\dfrac{A_{n}A_{m}+A_{n}%
\theta_{m}^{2}}{A_{m}^{2}}\left[  \boldsymbol{Z}\right]  _{in}^{2}\left[
\boldsymbol{Z}\right]  _{im}^{2}+\sum_{n=1}^{p}\sum_{m\neq n}\left[
\boldsymbol{Z}\right]  _{in}^{2}\left[  \boldsymbol{Z}\right]  _{im}^{2}\\
&  +2\sum_{n=1}^{p}\sum_{m\neq n}\dfrac{\theta_{n}\theta_{m}}{A_{m}}\left[
\boldsymbol{Z}\right]  _{in}^{3}\left[  \boldsymbol{Z}\right]  _{im}%
+\sum_{n=1}^{p}\sum_{m\neq n^{\prime},n^{\prime}\neq n,m\neq n}\dfrac
{A_{n}\theta_{m}\theta_{n^{\prime}}}{A_{m}A_{n^{\prime}}}\left[
\boldsymbol{Z}\right]  _{in}^{2}\left[  \boldsymbol{Z}\right]  _{im}\left[
\boldsymbol{Z}\right]  _{in^{\prime}}\\
&  =\sum_{n,m=1}^{p}\dfrac{A_{n}}{A_{m}}\left[  \boldsymbol{Z}\right]
_{in}^{2}\left[  \boldsymbol{Z}\right]  _{im}^{2}+\sum_{n,m=1}^{p}\left[
\boldsymbol{Z}\right]  _{in}^{2}\left[  \boldsymbol{Z}\right]  _{im}^{2}%
+\sum_{n,m,n^{\prime}=1}^{p}\dfrac{A_{n}\theta_{m}\theta_{n^{\prime}}}%
{A_{m}A_{n^{\prime}}}\left[  \boldsymbol{Z}\right]  _{in}^{2}\left[
\boldsymbol{Z}\right]  _{im}\left[  \boldsymbol{Z}\right]  _{in^{\prime}}.
\end{align*}
Using matrix notation, we can reexpress the above equation as%
\begin{align*}
\mathbb{E}\left(  \left[  \boldsymbol{\Xi}\right]  _{ii}^{2}\right)   &
=\left[  \boldsymbol{ZAZ}^{T}\right]  _{ii}\left[  \boldsymbol{ZA}%
^{-1}\boldsymbol{Z}^{T}\right]  _{ii}+\left[  \boldsymbol{ZZ}^{T}\right]
_{ii}^{2}+\left[  \boldsymbol{ZAZ}^{T}\right]  _{ii}\left[  \boldsymbol{ZA}%
^{-1}\boldsymbol{\theta}\right]  _{i}^{2}\\
&  \leq\mathrm{tr}\left(  \boldsymbol{ZAZ}^{T}\right)  \mathrm{tr}\left(
\boldsymbol{ZA}^{-1}\boldsymbol{Z}^{T}\right)  +\mathrm{tr}\left(
\boldsymbol{ZZ}^{T}\right)  ^{2}+\mathrm{tr}\left(  \boldsymbol{ZAZ}%
^{T}\right)  \mathrm{tr}\left(  \boldsymbol{\theta}^{T}\boldsymbol{A}%
^{-1}\boldsymbol{Z}^{T}\boldsymbol{ZA}^{-1}\boldsymbol{\theta}\right) \\
&  =\mathrm{tr}\left(  \boldsymbol{WXAX}^{T}\right)  \mathrm{tr}\left(
\boldsymbol{WXA}^{-1}\boldsymbol{X}^{T}\right)  +\mathrm{tr}\left(
\boldsymbol{WXX}^{T}\right)  ^{2}\\
&  +\mathrm{tr}\left(  \boldsymbol{WXAX}^{T}\right)  \mathrm{tr}\left(
\boldsymbol{\beta}^{T}\left(  \boldsymbol{XA}^{-1}\boldsymbol{X}^{T}\right)
\boldsymbol{W}\left(  \boldsymbol{XA}^{-1}\boldsymbol{X}^{T} \right)
\boldsymbol{\beta}\right)  ,
\end{align*}
which is $O\left(  p\right)  O\left(  p\right)  +O\left(  p\right)
^{2}+O\left(  p\right)  O\left(  p^{2}\right)  =O\left(  p^{3}\right)  $ by
conditions $\left(  \mathrm{D}\right)  $-$\left(  \mathrm{F}\right)  $. Note
also that condition $\left(  \mathrm{F}\right)  $ implies%
\[
d_{k}\leq\sum_{i=1}^{k}d_{i}=\mathrm{tr}\left(  \boldsymbol{W}^{-1/2}%
\boldsymbol{VW}^{-1/2}\right)  =\mathrm{tr}\left(  \boldsymbol{W}%
^{-1}\boldsymbol{V}\right)  =\mathrm{tr}\left(  \boldsymbol{W}^{-1}%
(\boldsymbol{XA}^{-1}\boldsymbol{X}^{T})^{-1}\right)  =O\left(  p^{-1}\right)
.
\]
Therefore, we have%
\[
\mathbb{E}\left(  d_{k}^{2}\left[  \boldsymbol{\Xi}\right]  _{ii}^{2}%
/p^{2}\right)  =O\left(  p^{-2}\right)  O\left(  p^{3}\right)  /p^{2}=O\left(
p^{-1}\right)  \rightarrow0,
\]
which proves
\[
\sup\limits_{0\leq\lambda\leq\infty}\left\vert \left(  \mathrm{II}\right)
\right\vert \rightarrow0\text{ in }L^{2},\ \ \ \ \text{as }p\rightarrow
\infty.
\]

To prove the uniform convergence of $\left(  \mathrm{III}\right)  _{1}$ to
zero in $L^{1}$, we note that%
\begin{align*}
\sup\limits_{\boldsymbol{\mu}\in\mathcal{L}}\left\vert \left(  \mathrm{III}%
\right)  _{1}\right\vert  &  =\dfrac{2}{p}\sup\limits_{\boldsymbol{\mu}%
\in\mathcal{L}}\left\vert \boldsymbol{\mu}^{T}\left(  \boldsymbol{Y}%
-\boldsymbol{\theta}\right)  \right\vert =\dfrac{2}{p}\sup
\limits_{\boldsymbol{\mu}\in\mathcal{L}}\left\vert \boldsymbol{\mu}%
^{T}\boldsymbol{P}_{\boldsymbol{X}}\left(  \boldsymbol{Y}-\boldsymbol{\theta
}\right)  \right\vert \\
&  \leq\dfrac{2}{p}\sup\limits_{\boldsymbol{\mu}\in\mathcal{L}}\left\Vert
\boldsymbol{\mu}\right\Vert \times\left\Vert \boldsymbol{P}_{\boldsymbol{X}%
}\left(  \boldsymbol{Y}-\boldsymbol{\theta}\right)  \right\Vert = \dfrac{2}%
{p}Mp^{\kappa}\left\Vert \boldsymbol{Y}\right\Vert \times\left\Vert
\boldsymbol{P}_{\boldsymbol{X}}\left(  \boldsymbol{Y}-\boldsymbol{\theta
}\right)  \right\Vert ,
\end{align*}
so by Cauchy-Schwarz inequality%
\begin{equation}
\mathbb{E}\left(  \sup\limits_{\boldsymbol{\mu}\in\mathcal{L}}\left\vert
\left(  \mathrm{III}\right)  _{1}\right\vert \right)  \leq2 M p^{\kappa
-1}\sqrt{\mathbb{E} \left(  \left\Vert \boldsymbol{Y}\right\Vert ^{2} \right)
}\sqrt{\mathbb{E}\left(  \left\Vert \boldsymbol{P}_{\boldsymbol{X}}\left(
\boldsymbol{Y}-\boldsymbol{\theta}\right)  \right\Vert ^{2}\right)  }.
\label{Cauchy-Schwarz Inequality for (III)_1 in Model II}%
\end{equation}
Under Model II, $\boldsymbol{\theta}=\boldsymbol{X}^{T}\boldsymbol{\beta}$, so
it follows that $\sum_{i=1}^{p}\theta_{i}^{2}=\left\Vert \boldsymbol{\theta
}\right\Vert ^{2}=\mathrm{tr}\left(  \boldsymbol{\beta\beta}^{T}%
\boldsymbol{XX}^{T}\right)  =O\left(  p\right)  $ by condition $\left(
\mathrm{E}\right)  $. Hence $\sqrt{\mathbb{E}\left(  \left\Vert \boldsymbol{Y}%
\right\Vert ^{2}\right)  }=\sqrt{\sum\nolimits_{i=1}^{p}\left(  \theta_{i}%
^{2}+A_{i}\right)  }=O\left(  p^{1/2}\right)  $. For the second term on the
right hand side of (\ref{Cauchy-Schwarz Inequality for (III)_1 in Model II}),
note that%
\begin{align*}
\mathbb{E}\left(  \left\Vert \boldsymbol{P}_{\boldsymbol{X}}\left(
\boldsymbol{Y}-\boldsymbol{\theta}\right)  \right\Vert ^{2}\right)   &
=\mathbb{E}\left(  \mathrm{tr}\left(  \boldsymbol{P}_{\boldsymbol{X}}\left(
\boldsymbol{Y}-\boldsymbol{\theta}\right)  \left(  \boldsymbol{Y}%
-\boldsymbol{\theta}\right)  ^{T}\right)  \right) \\
&  =\mathrm{tr}\left(  \boldsymbol{P}_{\boldsymbol{X}}\boldsymbol{A}\right)
=\mathrm{tr}\left(  \left(  \boldsymbol{XX}^{T}\right)  ^{-1}\boldsymbol{XAX}%
^{T}\right)  =O\left(  1\right)
\end{align*}
by conditions $\left(  \mathrm{D}\right)  $ and $\left(  \mathrm{E}\right)  $.
Thus, in aggregate, we have%
\[
\mathbb{E}\left(  \sup\limits_{\boldsymbol{\mu}\in\mathcal{L}}\left\vert
\left(  \mathrm{III}\right)  _{1}\right\vert \right)  \leq2Mp^{\kappa
-1}O\left(  p^{1/2}\right)  O\left(  1\right)  =o\left(  1\right)  .
\]

We finally consider the $\left(  \mathrm{III}\right)  _{2}$ term. We have%
\begin{align*}
\sup\limits_{0\leq\lambda\leq\infty,\;\boldsymbol{\mu}\in\mathcal{L}%
}\left\vert \left(  \mathrm{III}\right)  _{2}\right\vert  &  =\dfrac{2}{p}%
\sup\limits_{\boldsymbol{\mu}\in\mathcal{L}}\sup\limits_{0\leq\lambda
\leq\infty}\left\vert \sum_{i=1}^{k}\dfrac{\lambda d_{i}}{\lambda+d_{i}%
}\left[  \boldsymbol{ZA}^{-1}\boldsymbol{\mu}\left(  \boldsymbol{Y}%
-\boldsymbol{\theta}\right)  ^{T}\boldsymbol{Z}^{T}\right]  _{ii}\right\vert
\\
&  \leq\dfrac{2}{p}\sup\limits_{\boldsymbol{\mu}\in\mathcal{L}}\max
\limits_{1\leq j\leq k}\left\vert \sum_{i=j}^{k}d_{k}\left[  \boldsymbol{ZA}%
^{-1}\boldsymbol{\mu}\left(  \boldsymbol{Y}-\boldsymbol{\theta}\right)
^{T}\boldsymbol{Z}^{T}\right]  _{ii}\right\vert \\
&  \leq\dfrac{2d_{k}}{p}\sup\limits_{\boldsymbol{\mu}\in\mathcal{L}}\sum
_{i=1}^{k}\left\vert \left[  \boldsymbol{ZA}^{-1}\boldsymbol{\mu}\left(
\boldsymbol{Y}-\boldsymbol{\theta}\right)  ^{T}\boldsymbol{Z}^{T}\right]
_{ii}\right\vert \\
&  =\dfrac{2d_{k}}{p}\sup\limits_{\boldsymbol{\mu}\in\mathcal{L}}\sum
_{i=1}^{k}\left\vert \left[  \boldsymbol{ZA}^{-1}\boldsymbol{\mu}\right]
_{i}\left[  \boldsymbol{Z}\left(  \boldsymbol{Y}-\boldsymbol{\theta}\right)
\right]  _{i}\right\vert \\
&  \leq\dfrac{2d_{k}}{p}\sup\limits_{\boldsymbol{\mu}\in\mathcal{L}}\sqrt
{\sum_{i=1}^{k}\left[  \boldsymbol{ZA}^{-1}\boldsymbol{\mu}\right]  _{i}^{2}%
}\times\sqrt{\sum_{i=1}^{k}\left[  \boldsymbol{Z}\left(  \boldsymbol{Y}%
-\boldsymbol{\theta}\right)  \right]  _{i}^{2}}.
\end{align*}
Thus, by Cauchy-Schwarz inequality%
\[
\mathbb{E}\left(  \sup\limits_{0\leq\lambda\leq\infty,\;\boldsymbol{\mu}%
\in\mathcal{L}}\left\vert \left(  \mathrm{III}\right)  _{2}\right\vert
\right)  \leq\dfrac{2d_{k}}{p}\sqrt{\mathbb{E}\left(  \sup
\limits_{\boldsymbol{\mu}\in\mathcal{L}}\sum_{i=1}^{k}\left[  \boldsymbol{ZA}%
^{-1}\boldsymbol{\mu}\right]  _{i}^{2}\right)  } \times\sqrt{\mathbb{E}
\left(  \sum_{i=1}^{k}\left[  \boldsymbol{Z}\left(  \boldsymbol{Y}%
-\boldsymbol{\theta}\right)  \right]  _{i}^{2}\right)  }.
\]
Note that%
\begin{align*}
&  \sup\limits_{\boldsymbol{\mu}\in\mathcal{L}}\sum_{i=1}^{k}\left[
\boldsymbol{ZA}^{-1}\boldsymbol{\mu}\right]  _{i}^{2} =\sup
\limits_{\boldsymbol{\mu}\in\mathcal{L}}\sum_{i=1}^{k}\left(  \sum_{m=1}%
^{p}\left[  \boldsymbol{ZA}^{-1}\right]  _{im}\left[  \boldsymbol{\mu}\right]
_{m}\right)  ^{2}\\
&  \leq\sup\limits_{\boldsymbol{\mu}\in\mathcal{L}}\sum_{i=1}^{k}\left(
\sum_{m=1}^{p}\left[  \boldsymbol{ZA}^{-1}\right]  _{im}^{2}\times\sum
_{m=1}^{p}\left[  \boldsymbol{\mu}\right]  _{m}^{2}\right)  =\sup
\limits_{\boldsymbol{\mu}\in\mathcal{L}}\sum_{i=1}^{k}\left(  \left[
\boldsymbol{ZA}^{-2}\boldsymbol{Z}^{T}\right]  _{ii}\left\Vert \boldsymbol{\mu
}\right\Vert ^{2}\right) \\
&  =\mathrm{tr}\left(  \boldsymbol{ZA}^{-2}\boldsymbol{Z}^{T}\right)
\sup\limits_{\boldsymbol{\mu}\in\mathcal{L}}\left\Vert \boldsymbol{\mu
}\right\Vert ^{2} =\mathrm{tr}\left(  \boldsymbol{WXA}^{-2}\boldsymbol{X}%
^{T}\right)  \left(  Mp^{\kappa}\left\Vert \boldsymbol{Y}\right\Vert \right)
^{2}=o\left(  p^{2}\right)  \left\Vert \boldsymbol{Y}\right\Vert ^{2},
\end{align*}
where the last equality uses condition $\left(  \mathrm{G}\right)  $. Thus,%
\[
\mathbb{E}\left(  \sup\limits_{\boldsymbol{\mu}\in\mathcal{L}}\sum_{i=1}%
^{k}\left[  \boldsymbol{ZA}^{-1}\boldsymbol{\mu}\right]  _{i}^{2} \right)
=o\left(  p^{3}\right)  .
\]
Also note that%
\begin{align*}
\mathbb{E}\left(  \sum_{i=1}^{k}\left[  \boldsymbol{Z}\left(  \boldsymbol{Y}%
-\boldsymbol{\theta}\right)  \right]  _{i}^{2} \right)   &  =\mathbb{E}\left(
\mathrm{tr}\left(  \boldsymbol{Z}^{T}\boldsymbol{Z}\left(  \boldsymbol{Y}%
-\boldsymbol{\theta}\right)  \left(  \boldsymbol{Y}-\boldsymbol{\theta
}\right)  ^{T}\right)  \right) \\
&  =\mathrm{tr}\left(  \boldsymbol{Z}^{T}\boldsymbol{ZA}\right)
=\mathrm{tr}\left(  \boldsymbol{WXAX}^{T}\right)  =O\left(  p\right)
\end{align*}
by condition $\left(  \mathrm{D}\right)  $. Recall that $d_{k}=O\left(
p^{-1}\right)  $ by condition $\left(  \mathrm{F}\right)  $. It follows that%
\[
\mathbb{E}\left(  \sup\limits_{0\leq\lambda\leq\infty,\;\boldsymbol{\mu}%
\in\mathcal{L}}\left\vert \left(  \mathrm{III}\right)  _{2}\right\vert
\right)  \leq\dfrac{2}{p}O\left(  p^{-1}\right)  o\left(  p^{3/2}\right)
O\left(  p^{1/2}\right)  =o\left(  1\right)  ,
\]
which completes our proof.

\textbf{Proof of Lemma \ref{Lma.: WLS is in the restriction class}.} The fact
that $\boldsymbol{\hat{\mu}}^{\mathrm{OLS}}\in\mathcal{L}$ is trivial as
\[
\boldsymbol{\hat{\mu}}^{\mathrm{OLS}}=\boldsymbol{X}^{T}\left(
\boldsymbol{XX}^{T}\right)  ^{-1}\boldsymbol{XY}=\boldsymbol{P}%
_{\boldsymbol{X}}\boldsymbol{Y},
\]
while the projection matrix $\boldsymbol{P}_{\boldsymbol{X}}$ has induced
matrix $2$-norm $\left\Vert \boldsymbol{P}_{\boldsymbol{X}}\right\Vert _{2}%
=1$. Thus, $\left\Vert \boldsymbol{\hat{\mu}}^{\mathrm{OLS}}\right\Vert
\leq\left\Vert \boldsymbol{P}_{\boldsymbol{X}}\right\Vert _{2}\left\Vert
\boldsymbol{Y}\right\Vert =\left\Vert \boldsymbol{Y}\right\Vert $. For
$\boldsymbol{\hat{\mu}}^{\mathrm{WLS}}$, note that%
\begin{align*}
\boldsymbol{\hat{\mu}}^{\mathrm{WLS}}  &  =\boldsymbol{X}^{T}\left(
\boldsymbol{XA}^{-1}\boldsymbol{X}^{T}\right)  ^{-1}\boldsymbol{XA}%
^{-1}\boldsymbol{Y}\\
&  =\boldsymbol{A}^{1/2}\left(  \boldsymbol{XA}^{-1/2}\right)  ^{T}\left(
\boldsymbol{XA}^{-1/2}\left(  \boldsymbol{XA}^{-1/2}\right)  ^{T}\right)
^{-1}\left(  \boldsymbol{XA}^{-1/2}\right)  \boldsymbol{A}^{-1/2}%
\boldsymbol{Y}\\
&  =\boldsymbol{A}^{1/2}\left(  \boldsymbol{P}_{\boldsymbol{XA}^{-1/2}%
}\right)  \boldsymbol{A}^{-1/2}\boldsymbol{Y},
\end{align*}
where $\boldsymbol{P}_{\boldsymbol{XA}^{-1/2}}$ is the ordinary projection
matrix onto the row space of $\boldsymbol{XA}^{-1/2}$ and has induced matrix
$2$-norm $1$. It follows%
\[
\left\Vert \boldsymbol{\hat{\mu}}^{\mathrm{WLS}}\right\Vert \leq\left\Vert
\boldsymbol{A}^{1/2}\right\Vert _{2}\left\Vert \boldsymbol{P}_{\boldsymbol{A}%
^{-1/2}\boldsymbol{X}}\right\Vert _{2}\left\Vert \boldsymbol{A}^{-1/2}%
\right\Vert _{2}\left\Vert \boldsymbol{Y}\right\Vert =\max\limits_{1\leq i\leq
p}A_{i}^{1/2}\times\max\limits_{1\leq i\leq p}A_{i}^{-1/2}\times\left\Vert
\boldsymbol{Y}\right\Vert .
\]
Condition $\left(  \mathrm{A}\right)  $ gives%
\[
\max\limits_{1\leq i\leq p}A_{i}^{1/2}=(\max\limits_{1\leq i\leq p}A_{i}%
^{2})^{1/4}\leq(\sum_{i=1}^{p}A_{i}^{2})^{1/4}=O\left(  p^{1/4}\right)  .
\]
Similarly, condition $\left(  \mathrm{A}^{\prime}\right)  $ gives%
\[
\max\limits_{1\leq i\leq p}A_{i}^{-1/2}=(\max\limits_{1\leq i\leq p}%
A_{i}^{-2-\delta})^{1/\left(  4+2\delta\right)  }\leq(\sum_{i=1}^{p}%
A_{i}^{-2-\delta})^{1/\left(  4+2\delta\right)  }=O\left(  p^{1/\left(
4+2\delta\right)  }\right)  .
\]
We then have proved that%
\[
\left\Vert \boldsymbol{\hat{\mu}}^{\mathrm{WLS}}\right\Vert \leq O\left(
p^{1/4}\right)  O\left(  p^{1/\left(  4+2\delta\right)  }\right)  \left\Vert
\boldsymbol{Y}\right\Vert =O\left(  p^{\kappa}\right)  \left\Vert
\boldsymbol{Y}\right\Vert .
\]

\textbf{Proof of Theorem \ref{Thm.: OL Risk Optimality}.} To prove the first
assertion, note that%
\[
\mathrm{URE}\left(  \boldsymbol{\hat{B}}^{\mathrm{URE}},\boldsymbol{\hat{\mu}%
}^{\mathrm{URE}}\right)  \leq\mathrm{URE}\left(  \boldsymbol{\tilde{B}%
}^{\mathrm{OL}},\boldsymbol{\tilde{\mu}}^{\mathrm{OL}}\right)
\]
by the definition of $\boldsymbol{\hat{B}}^{\mathrm{URE}}$ and
$\boldsymbol{\hat{\mu}}^{\mathrm{URE}}$, so Theorem \ref{Thm.: Main Theorem}
implies that%
\begin{align}
&  l_{p}\left(  \boldsymbol{\theta},\boldsymbol{\hat{\theta}}^{\mathrm{URE}%
}\right)  -l_{p}\left(  \boldsymbol{\theta},\boldsymbol{\tilde{\theta}%
}^{\mathrm{OL}}\right) \nonumber\\
&  \leq l_{p}\left(  \boldsymbol{\theta},\boldsymbol{\hat{\theta}%
}^{\mathrm{URE}}\right)  -\mathrm{URE}\left(  \boldsymbol{\hat{B}%
}^{\mathrm{URE}},\boldsymbol{\hat{\mu}}^{\mathrm{URE}}\right)  +\mathrm{URE}%
\left(  \boldsymbol{\tilde{B}}^{\mathrm{OL}},\boldsymbol{\tilde{\mu}%
}^{\mathrm{OL}}\right)  -l_{p}\left(  \boldsymbol{\theta},\boldsymbol{\tilde
{\theta}}^{\mathrm{OL}}\right) \nonumber\\
&  \leq2\sup\limits_{\boldsymbol{B}\in\mathcal{B},\;\boldsymbol{\mu}%
\in\mathcal{L}}\left\vert \mathrm{URE}\left(  \boldsymbol{B},\boldsymbol{\mu
}\right)  -l_{p}\left(  \boldsymbol{\theta},\boldsymbol{\hat{\theta}%
}^{\boldsymbol{B},\boldsymbol{\mu}}\right)  \right\vert \underset{p\rightarrow
\infty}{\rightarrow}0\text{ in }L^{1}\text{ and in probability,}
\label{Loss diff. between URE and OL}%
\end{align}
where the second inequality uses the condition that $\boldsymbol{\hat{\mu}%
}^{\mathrm{URE}}\in\mathcal{L}$. Thus, for any $\epsilon>0$,%
\begin{align*}
&  \mathbb{P}\left(  l_{p}\left(  \boldsymbol{\theta},\boldsymbol{\hat{\theta
}}^{\mathrm{URE}}\right)  \geq l_{p}\left(  \boldsymbol{\theta}%
,\boldsymbol{\tilde{\theta}}^{\mathrm{OL}}\right)  +\epsilon\right) \\
&  \leq\mathbb{P}\left(  2\sup\limits_{\boldsymbol{B}\in\mathcal{B}%
,\;\boldsymbol{\mu}\in\mathcal{L}}\left\vert \mathrm{URE}\left(
\boldsymbol{B},\boldsymbol{\mu}\right)  -l_{p}\left(  \boldsymbol{\theta
},\boldsymbol{\hat{\theta}}^{\boldsymbol{B},\boldsymbol{\mu}}\right)
\right\vert \geq\epsilon\right)  \rightarrow0.
\end{align*}

To prove the second assertion, note that%
\[
l_{p}\left(  \boldsymbol{\theta},\boldsymbol{\tilde{\theta}}^{\mathrm{OL}%
}\right)  \leq l_{p}\left(  \boldsymbol{\theta},\boldsymbol{\hat{\theta}%
}^{\mathrm{URE}}\right)
\]
by the definition of $\boldsymbol{\tilde{\theta}}^{\mathrm{OL}}$ and the
condition $\boldsymbol{\hat{\mu}}^{\mathrm{URE}}\in\mathcal{L}$. Thus, taking
expectations on equation (\ref{Loss diff. between URE and OL}) easily gives
the second assertion.

\textbf{Proof of Corollary \ref{Cor.: Risk properties}.} Simply note that%
\[
l_{p}\left(  \boldsymbol{\theta},\boldsymbol{\tilde{\theta}}^{\mathrm{OL}%
}\right)  \leq l_{p}\left(  \boldsymbol{\theta},\boldsymbol{\hat{\theta}%
}^{\boldsymbol{\hat{B}}_{p},\boldsymbol{\hat{\mu}}_{p}}\right)
\]
by the definition of $\boldsymbol{\tilde{\theta}}^{\mathrm{OL}}$. Thus,%
\[
l_{p}\left(  \boldsymbol{\theta},\boldsymbol{\hat{\theta}}^{\mathrm{URE}%
}\right)  -l_{p}\left(  \boldsymbol{\theta},\boldsymbol{\hat{\theta}%
}^{\boldsymbol{\hat{B}}_{p},\boldsymbol{\hat{\mu}}_{p}}\right)  \leq
l_{p}\left(  \boldsymbol{\theta},\boldsymbol{\hat{\theta}}^{\mathrm{URE}%
}\right)  -l_{p}\left(  \boldsymbol{\theta},\boldsymbol{\tilde{\theta}%
}^{\mathrm{OL}}\right)  .
\]
Then Theorem \ref{Thm.: OL Risk Optimality} clearly implies the desired result.

\textbf{Proof of Theorem \ref{Thm.: Main Theorem for shrinking toward GLS}.}
We observe that%
\[
\mathrm{URE}_{\boldsymbol{M}}\left(  \boldsymbol{B}\right)  -l_{p}\left(
\boldsymbol{\theta},\boldsymbol{\hat{\theta}}^{\boldsymbol{B},\boldsymbol{\hat
{\mu}}^{\boldsymbol{M}}} \right)  =\mathrm{URE}\left(  \boldsymbol{B}%
,\boldsymbol{\hat{\mu}}^{\boldsymbol{M}}\right)  -l_{p}\left(
\boldsymbol{\theta},\boldsymbol{\hat{\theta}}^{\boldsymbol{B},\boldsymbol{\hat
{\mu}}^{\boldsymbol{M}}} \right)  +\dfrac{2}{p}\mathrm{tr}\left(
\boldsymbol{A}\left(  \boldsymbol{A}+\boldsymbol{B}\right)  ^{-1}%
\boldsymbol{P}_{\boldsymbol{M},\boldsymbol{X}}\boldsymbol{A}\right)  .
\]
Since%
\[
\sup\limits_{\boldsymbol{B}\in\mathcal{B}}\left\vert \mathrm{URE}\left(
\boldsymbol{B},\boldsymbol{\hat{\mu}}^{\boldsymbol{M}}\right)  -l_{p}\left(
\boldsymbol{\theta},\boldsymbol{\hat{\theta}}^{\boldsymbol{B},\boldsymbol{\hat
{\mu}}^{\boldsymbol{M}}} \right)  \right\vert \leq\sup\limits_{\boldsymbol{B}%
\in\mathcal{B},\mathcal{\;}\boldsymbol{\mu}\in\mathcal{L}}\left\vert
\mathrm{URE}\left(  \boldsymbol{B},\boldsymbol{\mu}\right)  -l_{p}\left(
\boldsymbol{\theta},\boldsymbol{\hat{\theta}}^{\boldsymbol{B},\boldsymbol{\mu
}}\right)  \right\vert \rightarrow0\text{ in }L^{1}%
\]
by Theorem \ref{Thm.: Main Theorem}, we only need to show that
\[
\sup\limits_{\boldsymbol{B}\in\mathcal{B}}\left\vert \dfrac{1}{p}%
\mathrm{tr}\left(  \boldsymbol{A}\left(  \boldsymbol{A}+\boldsymbol{B}\right)
^{-1}\boldsymbol{P}_{\boldsymbol{M},\boldsymbol{X}}\boldsymbol{A}\right)
\right\vert \rightarrow0\ \ \ \ \text{as }p\rightarrow\infty.
\]

Under Model I,
\begin{align*}
\mathrm{tr}\left(  \boldsymbol{A}\left(  \boldsymbol{A}+\boldsymbol{B}\right)
^{-1}\boldsymbol{P}_{\boldsymbol{M},\boldsymbol{X}}\boldsymbol{A}\right)   &
=\sum_{i=1}^{p}\frac{A_{i}}{A_{i}+\lambda}[\boldsymbol{P}_{\boldsymbol{M}%
,\boldsymbol{X}}\boldsymbol{A}]_{ii}\\
&  \leq\left(  \sum_{i=1}^{p}(\frac{A_{i}}{A_{i}+\lambda})^{2}\times\sum
_{i=1}^{p}[\boldsymbol{P}_{\boldsymbol{M},\boldsymbol{X}}\boldsymbol{A}%
]_{ii}^{2}\right)  ^{1/2}\\
&  \leq\left( p\times\sum_{i=1}^{p}\left[  \boldsymbol{P}_{\boldsymbol{M}%
,\boldsymbol{X}}\boldsymbol{A}\right]  _{ii}^{2}\right) ^{1/2}\\
&  \leq p^{1/2}\sqrt{\mathrm{tr}\left(  \boldsymbol{P}_{\boldsymbol{M}%
,\boldsymbol{X}}\boldsymbol{A}(\boldsymbol{P}_{\boldsymbol{M},\boldsymbol{X}%
}\boldsymbol{A})^{T}\right)  },\ \ \ \ \text{for all }\lambda\geq0,
\end{align*}
but $\mathrm{tr}\left(  \boldsymbol{P}_{\boldsymbol{M},\boldsymbol{X}%
}\boldsymbol{AAP}_{\boldsymbol{M},\boldsymbol{X}}^{T}\right)  =\mathrm{tr}%
\left(  \boldsymbol{X}^{T}\left(  \boldsymbol{XMX}^{T}\right)  ^{-1}%
\boldsymbol{XMA}^{2}\boldsymbol{MX}^{T}\left(  \boldsymbol{XMX}^{T}\right)
^{-1}\boldsymbol{X}\right)  $\newline$=\mathrm{tr}\left(  \left(
\boldsymbol{XMX}^{T}\right)  ^{-1}(\boldsymbol{XMA}^{2}\boldsymbol{MX}%
^{T})\left(  \boldsymbol{XMX}^{T}\right)  ^{-1}(\boldsymbol{XX}^{T})\right)
=O(1)$ by (\ref{conditionGLS}) and condition (E). Therefore,
\[
\sup\limits_{\boldsymbol{B}\in\mathcal{B}}\left\vert \dfrac{1}{p}%
\mathrm{tr}\left(  \boldsymbol{A}\left(  \boldsymbol{A}+\boldsymbol{B}\right)
^{-1}\boldsymbol{P}_{\boldsymbol{M},\boldsymbol{X}}\boldsymbol{A}\right)
\right\vert =\dfrac{1}{p}O\left(  p^{1/2}\right)  O(1)=O(p^{-1/2}%
)\rightarrow0.
\]

Under Model II, $\boldsymbol{A}\left(  \boldsymbol{A}+\boldsymbol{B}\right)
^{-1}=\boldsymbol{I}_{p}-\lambda\boldsymbol{Z}^{T}\left(  \lambda
\boldsymbol{I}_{k}+\boldsymbol{\Lambda}\right)  ^{-1}\boldsymbol{\Lambda
}\boldsymbol{Z}\boldsymbol{A}^{-1}$, where \newline$\boldsymbol{W}%
^{-1/2}\boldsymbol{VW}^{-1/2}=\boldsymbol{U\Lambda}\boldsymbol{U}^{T}$,
$\boldsymbol{\Lambda}=\mathrm{diag}\left(  d_{1},...,d_{k}\right)  $ with
$d_{1}\leq\cdots\leq d_{k}$, and $\boldsymbol{Z}=\boldsymbol{U}^{T}%
\boldsymbol{W}^{1/2}\boldsymbol{X}$ as defined in the proof of Theorem
\ref{Thm.: Main Theorem}. Thus,
\[
\mathrm{tr}\left(  \boldsymbol{A}\left(  \boldsymbol{A}+\boldsymbol{B}\right)
^{-1}\boldsymbol{P}_{\boldsymbol{M},\boldsymbol{X}}\boldsymbol{A}\right)
=\mathrm{tr}\left(  \boldsymbol{P}_{\boldsymbol{M},\boldsymbol{X}%
}\boldsymbol{A}\right)  -\mathrm{tr}\left(  \lambda\boldsymbol{Z}^{T}\left(
\lambda\boldsymbol{I}_{k}+\boldsymbol{\Lambda}\right)  ^{-1}%
\boldsymbol{\Lambda}\boldsymbol{Z}\boldsymbol{A}^{-1}\boldsymbol{P}%
_{\boldsymbol{M},\boldsymbol{X}}\boldsymbol{A}\right)  .
\]
We know that $\mathrm{tr}\left(  \boldsymbol{P}_{\boldsymbol{M},\boldsymbol{X}%
}\boldsymbol{A}\right)  =\mathrm{tr}\left(  \left(  \boldsymbol{XMX}%
^{T}\right)  ^{-1}(\boldsymbol{XMAX}^{T})\right)  =O(1)$ by the assumption
(\ref{conditionGLS}). $\mathrm{tr}\left(  \lambda\boldsymbol{Z}^{T}\left(
\lambda\boldsymbol{I}_{k}+\boldsymbol{\Lambda}\right)  ^{-1}%
\boldsymbol{\Lambda}\boldsymbol{Z}\boldsymbol{A}^{-1}\boldsymbol{P}%
_{\boldsymbol{M},\boldsymbol{X}}\boldsymbol{A}\right)  =\mathrm{tr}\left(
\lambda\left(  \lambda\boldsymbol{I}_{k}+\boldsymbol{\Lambda}\right)
^{-1}\boldsymbol{\Lambda}\boldsymbol{Z}\boldsymbol{A}^{-1}\boldsymbol{P}%
_{\boldsymbol{M},\boldsymbol{X}}\boldsymbol{AZ}^{T}\right)  $\newline%
$=\mathrm{tr}\left(  \lambda\left(  \lambda\boldsymbol{I}_{k}%
+\boldsymbol{\Lambda}\right)  ^{-1}\boldsymbol{\Lambda}\boldsymbol{Z}%
\boldsymbol{A}^{-1}\boldsymbol{X}^{T}\left(  \boldsymbol{XMX}^{T}\right)
^{-1}\boldsymbol{XMAZ}^{T}\right)  $. The Cauchy-Schwarz inequality for matrix
trace gives%
\begin{align*}
&  \left\vert \mathrm{tr}\left(  \left(  \lambda\left(  \lambda\boldsymbol{I}%
_{k}+\boldsymbol{\Lambda}\right)  ^{-1}\boldsymbol{\Lambda}\right)  \left(
\boldsymbol{ZA}^{-1}\boldsymbol{X}^{T}\left(  \boldsymbol{XMX}^{T}\right)
^{-1}\boldsymbol{XMAZ}^{T}\right)  \right)  \right\vert \\
&  \leq\mathrm{tr}^{1/2}\left(  (\lambda\left(  \lambda\boldsymbol{I}%
_{k}+\boldsymbol{\Lambda}\right)  ^{-1}\boldsymbol{\Lambda})^{2}\right) \\
&  \times\mathrm{tr}^{1/2}\left(  \boldsymbol{ZA}^{-1}\boldsymbol{X}%
^{T}\left(  \boldsymbol{XMX}^{T}\right)  ^{-1}\boldsymbol{XMAZ}^{T}%
\boldsymbol{ZAMX}^{T}\left(  \boldsymbol{XMX}^{T}\right)  ^{-1}\boldsymbol{XA}%
^{-1}\boldsymbol{Z}^{T}\right)  .
\end{align*}
Since
\begin{align*}
\mathrm{tr}\left(  (\lambda\left(  \lambda\boldsymbol{I}_{k}%
+\boldsymbol{\Lambda}\right)  ^{-1}\boldsymbol{\Lambda})^{2}\right)
=\sum_{i=1}^{k}\left( \dfrac{\lambda d_{i}}{\lambda+d_{i}} \right)^{2}\leq kd_{k}%
^{2}=O\left(  p^{-2}\right)\ \ \ \ \text{for all }\lambda\geq 0
\end{align*}
as shown in the
proof of Theorem \ref{Thm.: Main Theorem} and
\begin{align*}
&  \mathrm{tr}\left(  \boldsymbol{ZA}^{-1}\boldsymbol{X}^{T}\left(
\boldsymbol{XMX}^{T}\right)  ^{-1}\boldsymbol{XMAZ}^{T}\boldsymbol{ZAMX}%
^{T}\left(  \boldsymbol{XMX}^{T}\right)  ^{-1}\boldsymbol{XA}^{-1}%
\boldsymbol{Z}^{T}\right) \\
&  =\mathrm{tr}\left(  \left(  \boldsymbol{XMX}^{T}\right)  ^{-1}%
\boldsymbol{XMAZ}^{T}\boldsymbol{ZAMX}^{T}\left(  \boldsymbol{XMX}^{T}\right)
^{-1}\boldsymbol{XA}^{-1}\boldsymbol{Z}^{T}\boldsymbol{ZA}^{-1}\boldsymbol{X}%
^{T}\right) \\
&  =\mathrm{tr}\left(  \left(  \boldsymbol{XMX}^{T}\right)  ^{-1}%
(\boldsymbol{XMAX}^{T})\boldsymbol{W}(\boldsymbol{XAMX}^{T})\left(
\boldsymbol{XMX}^{T}\right)  ^{-1}(\boldsymbol{XA}^{-1}\boldsymbol{X}%
^{T})\boldsymbol{W}(\boldsymbol{XA}^{-1}\boldsymbol{X}^{T})\right) \\
&  =O(p^{2})
\end{align*}
from (\ref{conditionGLS}) and condition (F), we have
\[
\sup\limits_{\boldsymbol{B}\in\mathcal{B}}\left\vert \dfrac{1}{p}%
\mathrm{tr}\left(  \boldsymbol{A}\left(  \boldsymbol{A}+\boldsymbol{B}\right)
^{-1}\boldsymbol{P}_{\boldsymbol{M},\boldsymbol{X}}\boldsymbol{A}\right)
\right\vert =\dfrac{1}{p}\left(  O(1)+\sqrt{O\left(  p^{-2}\right)  \times
O(p^{2})}\right)  =O(p^{-1})\rightarrow0.
\]
This completes our proof of (\ref{unifconvGLS}). With this established, the
rest of the proof is identical to that of Theorem
\ref{Thm.: OL Risk Optimality} and Corollary \ref{Cor.: Risk properties}.

%
%
 \bibliographystyle{spmpsci}
 \bibliography{Sam,ju-chen}

\biblstarthook{References may be \textit{cited} in the text either by number (preferred) or by author/year.\footnote{Make sure that all references from the list are cited in the text. Those not cited should be moved to a separate \textit{Further Reading} section or chapter.} The reference list should ideally be \textit{sorted} in alphabetical order -- even if reference numbers are used for the their citation in the text. If there are several works by the same author, the following order should be used:
\begin{enumerate}
\item all works by the author alone, ordered chronologically by year of publication
\item all works by the author with a coauthor, ordered alphabetically by coauthor
\item all works by the author with several coauthors, ordered chronologically by year of publication.
\end{enumerate}
The \textit{styling} of references\footnote{Always use the standard abbreviation of a journal's name according to the ISSN \textit{List of Title Word Abbreviations}, see \url{http://www.issn.org/en/node/344}} depends on the subject of your book:
\begin{itemize}
\item The \textit{two} recommended styles for references in books on \textit{mathematical, physical, statistical and computer sciences} are depicted in ~\cite{science-contrib, science-online, science-mono, science-journal, science-DOI} and ~\cite{phys-online, phys-mono, phys-journal, phys-DOI, phys-contrib}.
\item Examples of the most commonly used reference style in books on \textit{Psychology, Social Sciences} are~\cite{psysoc-mono, psysoc-online,psysoc-journal, psysoc-contrib, psysoc-DOI}.
\item Examples for references in books on \textit{Humanities, Linguistics, Philosophy} are~\cite{humlinphil-journal, humlinphil-contrib, humlinphil-mono, humlinphil-online, humlinphil-DOI}.
\item Examples of the basic Springer style used in publications on a wide range of subjects such as \textit{Computer Science, Economics, Engineering, Geosciences, Life Sciences, Medicine, Biomedicine} are ~\cite{basic-contrib, basic-online, basic-journal, basic-DOI, basic-mono}.
\end{itemize}
}

\end{document}